\documentclass[11 pt]{article}%
\usepackage{bbm}
\usepackage[longnamesfirst]{natbib}
\usepackage{amssymb}
\usepackage{amsmath}
\usepackage{amsthm}
\usepackage{setspace}
\usepackage[colorlinks]{hyperref}
\usepackage{graphicx}
\usepackage{caption}
\usepackage{subcaption}
\usepackage{enumitem}
\usepackage[margin=1in]{geometry}
\usepackage{verbatim}
\usepackage{sectsty}
\usepackage[svgnames]{xcolor}
\usepackage{subfiles}
\usepackage{xargs}
\usepackage{titlesec}
\usepackage[colorinlistoftodos,textsize=tiny]{todonotes}
\usepackage[capitalize,nameinlink]{cleveref}
\usepackage[normalem]{ulem}
\usepackage[font={small,it},justification=justified]{caption}
\usepackage{xcolor}
\usepackage{pgf}
\usepackage{tikz}
\usepackage{amsfonts}%
\providecommand{\U}[1]{\protect\rule{.1in}{.1in}}
\setlist{noitemsep,parsep=6pt,partopsep=0pt,topsep=0pt}
\hypersetup{
pdftitle={Belief formation and the persistence of biased beliefs},
pdfauthor={Olivier Compte},
citecolor=DarkBlue,
bookmarksnumbered=true,
urlcolor=Indigo,linkcolor=DarkBlue
}
\theoremstyle{remark}

\theoremstyle{plain}

\renewcommand{\epsilon}{\varepsilon}
\usetikzlibrary{patterns, decorations.pathreplacing, arrows.meta,calc}
\let \savenumberline \numberline
\def \numberline#1{\savenumberline{#1.}}
\makeatletter
\renewcommand\@seccntformat[1]{\csname the#1\endcsname.{\hskip.7em\relax}}
\makeatother


\let\oldfootnote\footnote
\renewcommand\footnote[1]{\oldfootnote{\hspace{.5mm}#1}}
\titlespacing\section{0pt}{10pt plus 2pt minus 2pt}{4pt plus 2pt minus 2pt}
\titlespacing\subsection{0pt}{6pt plus 2pt minus 2pt}{2pt plus 2pt minus 2pt}
\titlespacing\subsubsection{0pt}{6pt plus 2pt minus 2pt}{0pt plus 2pt minus 2pt}
\titlespacing{\paragraph}{  0pt}{  0.5\baselineskip}{  1em}
\newcommand{\appendixref}[1]{\hyperref[#1]{Appendix \ref{#1}}}
\definecolor{dark-red}{rgb}{0.4,0.15,0.15}
\definecolor{dark-blue}{rgb}{0.15,0.15,0.75}
\definecolor{medium-blue}{rgb}{0,0,0.5}
\usetikzlibrary{shapes,arrows,automata,fit}
\tikzstyle{info}=[circle,thick,draw=black,fill=black!25,minimum size=4mm]
\tikzstyle{uninfo}=[circle,thick,draw=black,fill=white,minimum size=4mm]
\tikzstyle{inforecog}=[circle,line width=1mm,draw=black!50,fill=black!25,minimum size=4mm]
\tikzstyle{uninforecog}=[circle,line width=1mm,draw=black!50,fill=white,minimum size=4mm]
\tikzstyle{traded}=[draw, line width=1mm]
\tikzstyle{recog}=[draw=black!50, line width=1mm]
\sectionfont{\color{DarkRed}}
\subsectionfont{\color{DarkRed}}
\subsubsectionfont{\color{DarkRed}}
\newcommandx{\nageeb}[2][1=]{\todo[linecolor=blue,backgroundcolor=blue!25,bordercolor=blue,#1]{#2}}
\newcommandx{\andreas}[2][1=]{\todo[linecolor=black,backgroundcolor=black!25,bordercolor=black,#1]{#2}}
\newcommandx{\navin}[2][1=]{\todo[linecolor=red,backgroundcolor=red!25,bordercolor=red,#1]{#2}}
\begin{document}

\begin{titlepage}
\title{Belief formation and the persistence of biased beliefs}
\date{October 11$^{th}$ 2023}
\author{Olivier Compte\thanks{Affiliation: \textit{Paris School of Economics}, 48
Boulevard Jourdan, 75014 Paris and \textit{Ecole des Ponts Paris Tech},
\href{mailto:olivier.compte@gmail.com}{\color{dark-blue}olivier.compte@gmail.com}.  This paper revisits \textquotedblleft Mental Processes and Decision
Making\textquotedblright \ by Olivier Compte and Andrew Postewaite (\cite{compte10}) }}
\maketitle
\begin{abstract}
\noindent
We propose a belief-formation model where agents attempt to discriminate between two theories,
and where the asymmetry in strength between confirming and disconfirming evidence tilts beliefs
in favor of theories that generate strong (and possibly rare) confirming evidence and weak (and frequent)
disconfirming evidence. In our model, limitations on information processing provide incentives to censor weak
evidence, with the consequence that for some discrimination problems, evidence may become mostly one-sided, independently of the true underlying theory.
Sophisticated agents who know the characteristics of the censored data-generating process are not lured by this accumulation of ``evidence'', but less sophisticated ones end up with biased beliefs.
Keywords: Artificial Intelligence, Learning, Bounded Rationality, Biased Beliefs, Bayesian Models
JEL Classification Codes: C63, C73, D43, D83, D9, L51
\end{abstract}
\thispagestyle{empty}
\end{titlepage}

\tableofcontents

\setstretch{1.1}

\clearpage
\setcounter{tocdepth}{2} \thispagestyle{empty} \clearpage
\setcounter{page}{1}

\onehalfspacing

\section{Introduction}

How do people aggregate multiple pieces of information? How do people form
beliefs? Why would beliefs remain biased in favor of a particular theory even
as disconfirming evidence accumulates?

In economics, these questions are generally addressed through a Bayesian lens:
considering agents attempting to discriminate between, say, two states of the
world and getting signals imperfectly correlated with the underlying state of
the world, we generally assume agents form beliefs using Bayes rule. This rule
requires using the precise characteristics of the discrimination problem
considered (i.e., the joint distribution over states and signals), and
incorrect beliefs arise when agents hold incorrect priors about the
distribution over states of the world, or incorrect/misspecified priors about
the signal-generating process (i.e., the conditional distribution over signals
given states). Incorrect beliefs may also arise when beliefs directly enter
preferences or have instrumental value, as then, there is a direct motive for
holding biased beliefs, despite the suboptimal choices these distortion may generate.

We propose an alternative explanation based on the asymmetry in strength
between confirming and disconfirming evidence. In our model, agents process
\textit{coarse perceptions} of the signals they get (i.e., whether the
evidence is confirming one state of the world or the other), and then
aggregate these perceptions through a simple mental system having a
\textit{limited number of mental states}. These limitations on information
processing provide incentives to censor weak evidence, with the consequence
that for \textit{some} discrimination problems (those for which the evidence
confirming a particular state of the world is strong and rare, while
disconfirming evidence is frequent but weak), evidence may become mostly
one-sided (and confirming), independently of the underlying state. We argue
that many superstitions or folk beliefs, such as lunar effects, share this
asymmetric-strength structure, which, we also argue, is conducive to biased beliefs.

This last piece requires further explanation, as for a Bayesian, getting
one-sided evidence \textit{independently of the underlying state} should mean
that the ``evidence" is not informative after all, and thus she should not be
lured by the ``evidence" and stick to her priors.

We obtain biased beliefs because, we assume, belief formation cannot be tuned
to the precise characteristic of each discrimination problem considered, but
only on average over a range of discrimination problems. Specifically, we
depart from the classic Bayesian path in two ways: (i) in the spirit of the
algorithmic literature, we envision belief-formation as an all-purpose rule or
algorithm that applies \textit{across many }(information aggregation)\textit{
problems};\footnote{At an abstract level, Bayesian updating can also be viewed
as algorithm that produces beliefs. But it does so using precise
characteristics of the data-generating process. By all-purpose, we mean that
it does not use precise characteristics of the data-generating process.} (ii)
we assume that, \textit{to some limited extent}, signal processing and the
belief-formation rule evolve in ways that improve the agent's welfare,
\textit{on average over the information aggregation problems faced.}

Said differently, our approach allows the agent to tune belief formation in
the direction of welfare improvements. So in that sense we depart from the
literature that assumes Bayes rule using exogenously given misspecifications.
But we also assume that this tuning cannot be problem-specific, implying that
for some problems, posterior beliefs will likely depart from the
(problem-specific) Bayesian ones. Our contribution is in identifying the kind
of information structure that give rise to systematic errors, and the kind of
errors that agents fall pray to.

Furthermore, we hope this paper will be seen as a (constructive) critique of
the Bayesian methodology echoing well-known ones (\cite{wilson87}), whereby
agents' behavior end up being tuned to modelling features that seem outside
the scope of what they can reasonably apprehend. In our model, agents process
perceptions correlated with the underlying state, where perceptions are meant
to capture the agent's correct (though coarse) understanding of the signals'
informational content. These perceptions are then eventually aggregated into a
mental state. Understanding the informational content of a signal is one
thing. Understanding the process that generates perceptions (and, further,
mental states) conditional on each possible state of the world is another
matter. Our approach allows \textit{problem-specific perceptions}, but it also
rules out problem-specific optimization of belief-formation rules, preventing
the fine-tuning of posterior beliefs to the particular process that generates
these perceptions.

\subsection{Some classic explanations for superstitions and other folk
beliefs}

Folk beliefs often have the structure of a particular circumstance (C) or act
increasing the chance of an otherwise rare event (E); a sort of illusory
correlation (\cite{chapman67}) between C and E, where one overestimates the
frequency of occurrences of the sequence C-E.

A common explanation for the existence of such biases is that looking for
patterns in the environment has fitness value -- predicting the future or the
imminence of danger is useful,\footnote{See \cite{beck07}} and if the cost of
holding erroneous beliefs is small compared to the potential benefits, taking
the Pascalian bet is a good option: why not drink the miraculous water or
repent if this has the slightest chance of curing illness. In essence, the
explanation is based on the idea that beliefs are inevitably incorrect to some
extent and that some errors are less costly than others.

Still, one could be surprised that erroneous beliefs persist even (and
sometimes even more so) among people that are repeatedly confronted with
disconfirming evidence. It is not uncommon for nurses working in maternity
wards to believe in lunar effects (\cite{abell79}), for example the fact that
a full moon would increase the number of (unprogrammed) baby deliveries. Or at
the very least, these erroneous beliefs seem inconsistent with Bayesian
modelling, where eventually, after being exposed to data for long enough,
correct beliefs should prevail.

Outside the Bayesian sphere, one plausible explanation for some biased beliefs
is that they have \textit{instrumental value}: some biases may have a direct
positive effect on well-being or performance either because they reduce
anxiety, improve focus or give a sense of control. This includes many
(personal) superstitions such as the protection from Bad Luck conferred by
charms or amulets,\footnote{See for example \cite{hildburgh51}, who suggests
that amulets act an anxiety reducer, which fosters good lactation.} or the
powers conferred by magical thoughts and other ritualized or routine
behaviors.\footnote{This also includes placebo effects: an inactive treatment
may have positive health effect, so long as you believe it does.} Holding such
beliefs generates direct (first-order) gains and, if not excessively biased --
magic thoughts giving a sense of invincibility are potentially harmful, only
second-order losses.\footnote{This trade-off is for example examined in
\cite{compte04}, where biased beliefs about chances of success positively
affect performance. See \cite{kosegi06} for the case where beliefs directly
affect preferences. See also \cite{brunnermeier05}.}

Another plausible explanation for biased beliefs is the \textit{confirmation
bias}: once the seed of a belief is planted in people's mind, this belief
tends to persist even when erroneous because evidence is then processed with a
bias; people are more likely to see/look for/process evidence confirming the
belief, rather than disconfirming it.\footnote{The negative consequences of
the confirmation bias is clear (\cite{rabin99}). The possible fitness value of
the confirmation bias is discussed in models where agents lack will-power
(\cite{benabou02,benabou04}), modelled as a discrepancy between the welfare
criterion and the decision rule. Plausibly however, in the same way that some
biases in beliefs contribute to reduce anxiety, there could be some reassuring
value to seeing one's beliefs confirmed, a reassurance that has a first-order
effect on welfare in the same way that confidence does. The fitness value of
the confirmation bias has also been discussed within the perspective of social
interactions (see \cite{peters20}).}

Still, some beliefs seem more easily confirmed than others: if one starts with
the belief that the full moon has \textit{no effect} on baby deliveries, how
strongly will that belief be reinforced by the observation of hospital tension
on a non-full moon day? Or at least, for lunar effects, providing evidence in
favor of a lunar effect seems easier than providing evidence against it. A
single coincidence of a full moon and a high number of deliveries seem to be
strong evidence in favor of the theory, which cannot be matched in strength by
a single instance of a high number of deliveries without full moon: these
kinds of bad days just happens.

This asymmetry between the strengths of confirming and disconfirming evidence
is at the heart of our argument: we shall argue that beliefs are easily tilted
in favor of theories that generate strong (and possibly rare) confirming
evidence and weak (and frequent) disconfirming evidence, even when these
theories are untrue.


\subsection{Main modelling assumptions.}

We consider a family of decision problems over two alternatives $1$ and $2$
where many signals are processed prior to decision making. There are two
underlying states $\theta=1$,$2$, defining which alternative is the better
one, and signals potentially permit the agent to discriminate between the two
underlying states. Our model has four main ingredients:\textit{\ }

(a)\textit{\ A coarse mental system}: the agent gets signals that may vary in
informativeness, but she only processes the direction of the evidence (rather
than its strength), aggregating multiple evidence through a simple mental
system a la Wilson, with an odd number of mental states
$S=\{-K,...,-1,0,1,..K\}$. Specifically, starting from $s=0$, each signal
received may either be processed as confirming $\theta=1$ (this event is
denoted $\widetilde{\theta}=1$), generating a one-step move to the right (if
possible), or confirming $\theta=2$ (this event is denoted $\widetilde{\theta
}=2$), generating a one-step move to the left (if possible), or uninformative
($\widetilde{\theta}=0$), generating no move. With $K=2$, we have:

\begin{figure}[h]
\centering
\includegraphics[scale=0.6]{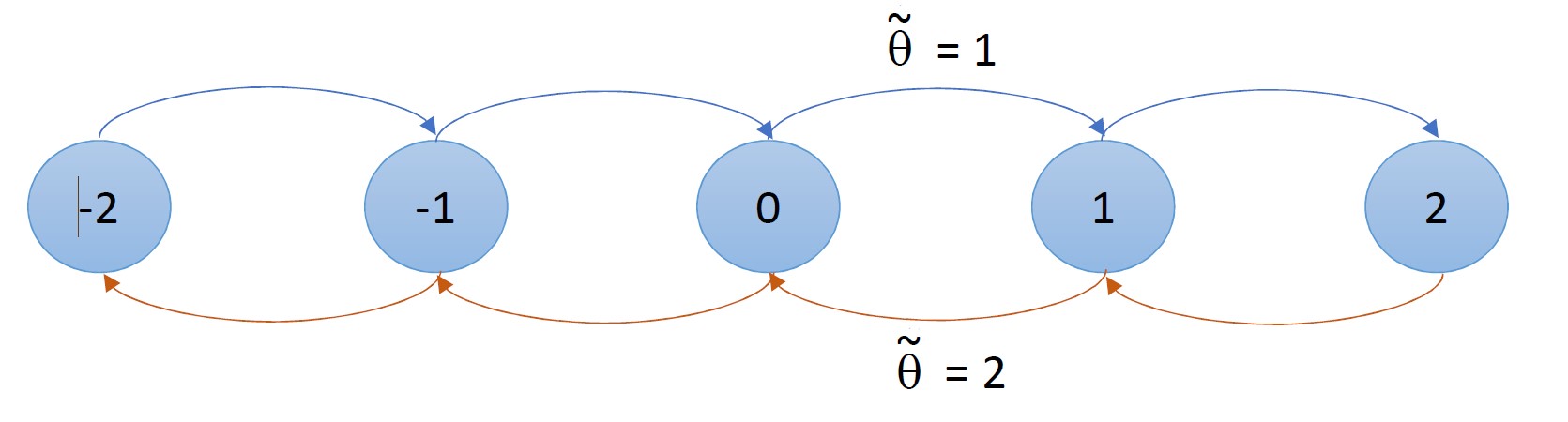}\caption{Coarse processing with $K=2$}%
\label{f0}%
\end{figure}

(b) \textit{limitations on how beliefs are formed: }each agent is equipped
with a possibly noisy prior, denoted $\widetilde{\rho},$ and we postulate an
``all-purpose" family of belief-formation strategies mapping priors and mental
state to a posterior belief.\ Specifically, expressing beliefs using
likelihood of 1 vs.~2, we assume that the agent's posterior belief in mental
state $s$ is
\begin{equation}
\widetilde{\rho}d^{s} \tag{P}\label{P}%
\end{equation}
where $d$ is a parameter that characterizes the degree to which the agent's
mental state affects posterior beliefs, or the\textit{\ discriminatory}
\textit{power} of the belief-formation rule. Given this (subjective) posterior
belief, the agent takes a decision that\ (subjectively) maximizes welfare.

(c)\textit{\ An option to censor weak evidence}. This option allows the agent
to focus on the more informative signals: only strong enough signals trigger
changes in the mental state, with a threshold strength parameterized by a
scalar $\beta$.\footnote{Formally, we shall say that a signal $x$ confirms (or
is evidence for) $\theta$, if signal $x$ is more likely under $\theta$ than
under $\theta^{\prime}\neq\theta)$. The strength of the evidence is then
defined as the ratio of probabilities of receiving $x$ under $\theta$ and
under $\theta^{\prime}$.} Technically, raising $\beta$ may affect the
distribution over $\widetilde{\theta}$ given $\theta$, hence the distribution
over mental states (hence beliefs). \smallskip

(d)\textit{\ Ex ante optimization of censoring and power. }The agent can
adjust the censoring $\beta$ and the power $d$ so as to maximize welfare, but
the optimization is done (ex ante) on average across problems, not contingent
of each discrimination problem faced.\footnote{This is the sense in which the
family (\ref{P}) is ``all-purpose". Technically, this means that $\beta$ and
$d$ are adjusted at an ex ante stage, before the signal-generating process is
selected.}

\subsection{Main intuitions.\textbf{\ }}

\textit{Regular and irregular problems.} Given the coarse processing (a)
assumed, the relevant characteristics of the signal-generating process will
reduce to the conditional probabilities $p_{\widetilde{\theta}\theta}$ of
processing evidence for state $\widetilde{\theta}\in\{1,2\}$ when the state is
$\theta$ (conditional on processing evidence)\footnote{That is, conditional on
processing evidence, we define $p_{\widetilde{\theta}\theta}$ as the
probability of processing evidence in favor of $\widetilde{\theta}$ when the
underlying state is $\theta$, so $p_{1\theta}+p_{2\theta}=1$.} and we shall
say that a problem is \textit{regular} iff%
\[
p_{\theta\theta}>1/2\text{ for each }\theta
\]
Intuitively, this means that for a regular problem, the agent's mental state
leans to the right when the underlying state is $1$, and to the left when the
underlying state is $2$. Discrimination between the two underlying states is
thus relatively easy. If, across all the problems faced, regular problems are
preponderant, then the individual has incentives to set the discriminatory
power parameter $d$ above 1 (because indeed the mental system is truly informative).

However, if there are problems for which, given censoring, the probabilities
$p_{\theta\theta}$ satisfy
\[
p_{11}>1/2\text{ and }p_{22}<1/2,
\]
the individual's mental state will lean to the right \textit{independently of
the underlying state}, and he will thus erroneously end up with beliefs
favoring theory 1 even in events where $\theta=2$: in these cases, processing
information moves posteriors \textit{away from the truth} and possibly
deteriorates welfare. For such problems, the agent would have been better off
setting $d=1$. The agent's inability to adjust $d$ to the characteristics
$p_{\widetilde{\theta}\theta}$ of the problem considered will be
key.\smallskip

\textit{Censoring.} In essence, at the margin, weak evidence adds noise to the
mental system, generating moves to the right and left with almost equal
probability. Censoring weak evidence eliminates these noisy moves. Does this
enhance welfare?

For a Bayesian who knows the signal generating process, the answer is positive
in most cases (though not all cases), and the reason is that mental states
being a scarce resource, one is generally better off limiting the use of
mental-state changes to sufficiently informative signals.\footnote{One
intuition is that given the limited number of mental states, posterior
Bayesian beliefs differ substantially from one another (across mental states).
Changing state after a poorly informative\ signal triggers a change in
posterior that seems unjustified. In most cases, avoiding these unjustified
changes is welfare increasing. In some rare cases however, for example when an
extreme mental state is very likely under both $\theta$, adding noise may
increase the informativeness of the mental system, as we further explain in
Section~\ref{subsectionBayesianCaseWeakEvidence}.}

For our less sophisticated agent, the answer depends on whether the problem is
regular or not. For regular problems, censoring weak evidence induces an
increase in both $p_{11}$ and $p_{22}$:
the correlation between the underlying state and the mental state is improved
and the discrimination between underlying states is improved --~and more
mental states help.

For irregular problems, with beliefs pointing towards, say $\widehat{\theta}$,
irrespective of the underlying state, the effect is opposite, reinforcing the
trend towards $\widehat{\theta}$: the balance between evidence confirming and
disconfirming $\widehat{\theta}$ becomes more favorable to $\widehat{\theta}$
independently of the underlying state, and when $\widehat{\theta}\neq\theta$,
this is potentially harmful for welfare (and even more so when there are more
mental states).

Optimal censoring of evidence trades off the two effects above, and to the
extent that regular problems are preponderant on average, the decision maker
has incentives to censor weak evidence.\smallskip

\textit{Superstition-prone problems.} The last piece of our argument consists
in observing that censoring weak evidence affects the type of problems that
are irregular as well as the underlying state that gets most likely confirmed:
problems for which a state mostly generate weak confirming evidence become
highly irregular when this weak evidence is censored. These types of problems
are thus prone to superstitious beliefs. For example, in\ evaluating whether a
rare circumstance $C$ has a positive influence on the probability that a rare
event $E$ occurs ($\theta=1$), or no influence ($\theta=2$), the only event
delivering strong evidence is $C-E$, and it favors $\theta=1$.\footnote{Since
$E$ is rare, $C-\overline{E}$ cannot be very informative, and when $C$ is rare
too, $\theta$ cannot affect much the occurence of $\overline{C}-E$.}\smallskip

\textit{Framing and pooling}. Finally, we use our framework to discuss the
importance of framing (i.e., which alternative theories $\theta$ are compared)
and pooling (i.e., how information or signals are structured) in fostering
biased beliefs. For a Bayesian, neither framing nor pooling alter the
direction of learning: beliefs on average lean towards the truth. For our less
sophisticated agent, both framing and pooling may affect which signals remain
strong enough evidence and get processed, possibly pushing beliefs away from
the truth.\footnote{This includes censoring, that is, pooling some a priori
informative signals with the many instances where signals are absent. For a
Bayesian, this ``missing data" event would become informative, while for our
agent, it would typically be too weak to be processed.}$^{,}$\footnote{For
example, assume that $C$ is not rare but $\overline{C}-E$ and $\overline
{C}-\overline{E}$ are pooled. Then the only potentially discriminating events
are $C-E$ and $C-\overline{E}$. Based on these events only, a Bayesian's
belief would lean towards the truth, while if $E$ is rare enough,beliefs of
our agent censoring weak evidence would lean towards $\theta=1$ independently
of the underlying state.}

In a similar vein, we discuss the effect of processing signals in batches
rather than sequentially as they arrive. With large enough batches, problems
become regular, so infrequent processing likely reduce biases.\smallskip

In summary, while the incentives to ignore weak evidence seem unavoidable, we
conclude that some problems are more prone to superstitions than others
because of asymmetries in strength of evidence for and against them. This
being said, people with a better understanding of the inherent biases of the
data generating process will be less prey to these biased beliefs, exerting
some form of skepticism, either by reducing the number of mental states, or,
when stakes are high, attributing less power to their mental system.

\subsection{\textbf{Discussion of modeling assumptions. }}

Our assumptions regarding belief formation depart from typical decision models
in several ways.

First, we attempt to model agents who form beliefs without much knowledge of
the process that generates \textit{perceptions} or eventually \textit{mental
states.}
The classic Bayesian route would remain a technically feasible modelling
option, but this route would involve simultaneous learning about the
underlying state and the data-generating process, hence would be cognitively
demanding for the agent and challenging for the analyst. Rather than following
this Bayesian route, we assume that the agent adopts a \textit{simple}
belief-formation rule that applies across discrimination
problems.\footnote{The belief-formation rules considered are simple enough
that endogenous adaptation to the particular characteristics $p$ of the
perception-generating process is not possible.}

Second, we endogenize signal processing and belief formation, in the sense
that, to a limited extent, we allow these to adjust to the economic
environment (in the direction of welfare improvement). That is, we do not take
for granted that agents would use Bayes rule or any other exogenously given
rule. We rather think of signal processing and belief formation as the result
of an evolutionary process that selects rules that enhance welfare. Said
differently, signal processing and belief formation jointly define \textit{a
strategy} from signals to posterior beliefs, and we assume the agent adopts a
strategy that maximizes welfare (ex ante), within a limited family of such
strategies (parameterized by a level of censoring $\beta$ and a discrimination
power $d$).

The particular family restriction (\ref{P})$\ $has been chosen for pedagogical
reasons: first it coincides with Bayesian updating in some special cases where
the data generating process is known, second it offers a simple way to
characterize the influence of mental processing on posterior beliefs. In
addition, many of the insights presented in this paper, including the
incentives to censor weak evidence, do not depend on the particular family
chosen, but just on the fact that beliefs are monotone in the mental state $s$.

Finally, note that we do not model how the adjustments of $\beta$ and $d$ are
made, though reinforcement learning or evolution is a natural candidate. In
that respect, we follow one of the classic route in game theory which keeps
unmodelled how players come up best responses. In any event, learning which
strategy is optimal within a simple family of strategies is certainly easier
than if no restrictions were put on the set of feasible belief-formation
strategies.\footnote{This feature of the model is inspired from
\cite{compte18}, which is more generally concerned with modelling agents
dealing with complex environment, and which uses strategy restrictions as a
modelling device to ensure that optimal behavior is not too finely tuned to
modelling details that agents cannot plausibly know.}

\subsection{Related work. \ }

We mentioned the instrumental value of beliefs and the confirmation bias as
two plausible explanations for the persistence of superstitions. Our
explanation is not inconsistent with these. We argue that some discrimination
problems are more prone to superstitions than others, which also implies that
for these problems, some instrumental value of superstitions, if present, will
be more easily derived (or the confirmation bias more easily sustained).

Another explanation for superstitious beliefs is due to \cite{chapman67}, who
coined the term \textquotedblleft illusory correlation". Chapman and Chapman
run an experiment in which subjects are presented associations of two words
(sequentially), and then later asked about the most frequent pairs. Among
these pairs, subjects tend to overweight the presence of \textquotedblleft
natural associations\textquotedblright\ such as \textquotedblleft lion-tiger".
Overweighting these \textquotedblleft natural associations" may bias the
judgment about the existence of correlations in the data. \cite{tversky1973}
see this an example of the availability heuristics. The \textquotedblleft
lion-tiger" pair being more natural, it is readily available in the brain and
becomes over-weighted when one tries to estimate ex post its occurrences in
the data (consisting in a list of paired words).

In a similar vein, one could argue that \textquotedblleft moon affecting
deliveries" is a natural association (the moon affects tides, why not a
woman's womb), and that as a result these events get over-represented in
people's mind. We provide an informativeness-based story for this
over-representation: the conjunction \textquotedblleft full-moon and many
deliveries" is more easily recorded or recalled than other events because of
an \textit{informativeness asymmetry}.

From a theory perspective, our paper is related to Robert Wilson's critique
(\cite{wilson87}), who argues that economic theories or mechanisms build on
potentially fragile ground, with optimal mechanisms tuned to details of the
economic environment that the mechanism designer cannot plausibly know. A
similar critique holds for agents finely adjusting strategies to details of a
model they cannot plausibly know. We address this critique by keeping the
number of strategic instruments limited ($\beta$ and $d$ are the only two
instruments), which effectively prevents the agent from adjusting its
belief-formation strategy to each particular perception-generating process.

Our model itself is closest to \cite{awilson14}'s work (as well as
\cite{compte10}), with an agent choosing an action after receiving a (random)
number of signals. The issue in \cite{awilson14} is the optimal use of a
limited number of states, which includes the optimal design of transition
probabilities between states, conditional on the signal received. When a long
sequence of signals is available (as in \cite{cover70}), the optimal use of
signals consists in organizing moves as in Figure 1, focusing only on the most
informative signal confirming $\theta=1$ (for moves to the right) or
$\theta=2$ (for moves to the left), and dealing with the asymmetry in strength
of evidence by adjusting the probability of moving away from an extreme state.
In that model, weak evidence is thus ignored (though what is considered weak
depends on the direction of evidence), and the asymmetry between the
frequencies of moves to the right and left are corrected by an appropriate
choice of transition probabilities at extreme states. Both of these features
(i.e. contingent censoring and contingent moves at extreme states) rely on
precise knowledge of the distribution over signals, which we do not assume.

The role played in our model by signals of asymmetric informational strength
echoes some insights of the \textit{mental accounting literature}. In
comparing two alternatives $A$ and $B$, agents may need to process many
signals related to the benefits or drawback of taking $A$ over $B$. For
example, if each alternative has many dimensions, aspects or attributes,
comparisons can be made on each dimension, each providing a potentially useful
signal. As noted early on by \cite{tversky69}, decision anomalies (such as
intransitivity of choices) may arise when the agent mostly focuses on the
dimensions where differences are more striking, or ignore dimensions where
differences are small compared to differences in other dimensions.\footnote{In
the same vein, \cite{rubinstein88} formalizes a notion of similarity providing
an explanation of the Allais Paradox. \cite{bordalo12,bordalo13} formalizes a
notion of salience explaining context dependent choices. \cite{koszegi12}
provides a general model of focus defining how, given the choice set, the
range of utility variations on each dimension affects which dimensions get
more heavily weighted. See also \cite{gabaix14} who limits the number of
dimensions that get attention (through processing costs or penalties in the
spirit of \cite{tibshirani96}), and endogenizes which ones get it.} More
generally, when few, say, positive gains are compared with numerous yet small
losses that each seem negligible and eventually ignored, this tilts the
decision in favor of the one yielding the large positive gains. While the
ability to detect or notice differences certainly matters in affecting the
degree to which some signals or dimensions are ignored, our model suggests
that there is an incentive to ignore weak signals that may go beyond this
(psychometrics-inspired) technological constraint.

In our model, signals are classified into classes: a signal is perceived as
evidence for $\theta$ if the likelihood that it has been generated by $\theta$
(as opposed to the other underlying state) is sufficiently large (by a factor
$1+\beta$). One interpretation is that the agent's mental state move only if
the signal's informational content is sufficiently strong. This likelihood
ratio rule can be seen as a way to evaluate the goodness of fit between two
alternative models, and it has been explored (assuming $\beta=0$) by
\cite{schwartzstein21} to model whether a sender (strategically proposing some
theory $\theta^{\prime}$) can persuade a receiver (initially holding some
theory $\theta$) to change her state of mind (and adopt $\theta^{\prime}$).


Finally, we contrast our work with the literature that explain biases through
agents forming beliefs based on a misspecified (or incomplete) model of the
environment (See \cite{esponda16} and \cite{spiegler16}, for example).

The paper is organized as follows. We present the model in Section 2. We
analyze optimal belief formation in Section 3. Next we examine incentives to
ignore weak evidence. In Section 5 we discuss how signal-strength asymmetries
may lead to persistently biased beliefs. We also provide examples of problems
where such asymmetries arise, also explaining why pooling of signals and
framing may modify these asymmetries. In Section 6 we discuss various
extensions of the model (fewer signals, more mental states, more underlying states)

\section{The model}

\subsection{Preferences and uncertainty}

We consider a family of decision problems, each having the following
structure: there are two possible \textit{states of the world} $\theta=1,2$
and after processing a sequence of signals, the agent eventually chooses
between two\textit{\ alternatives}, $a\in A=\{1,2\}$. Each problem has a
\textit{specific} payoff structure and a \textit{specific} signal structure.

The payoff structure is characterized by a \textit{payoff matrix}, where
$g(a,\theta)$ is the payoff to the agent when she takes action $a$ in state
$\theta$:%
\[%
\begin{tabular}
[c]{c|cc}%
$g(a,\theta)$ & $1$ & $2$\\\hline
$1$ & $1-\gamma$ & $0$\\\hline
$2$ & $0$ & $\gamma$%
\end{tabular}
\ \
\]
where $\gamma\in\lbrack0,1]$. When $\gamma>1/2$, taking the right decision is
more important when the state is 2 than where the state is 1. The ratio
$\Gamma=\gamma/(1-\gamma)$ characterizes that relative importance.

We assume that $\theta=1$ for a fraction $\pi$ of the problems, independently
of the payoff characteristic $\gamma$. The fraction $\pi$ thus characterizes
some \textit{objective uncertainty} about the true state, and we let $\rho=$
$\pi/(1-\pi)$ denote the odds ratio. This objective uncertainty does not
necessarily coincide with the agent's \textit{initial/prior belief}: we allow
for some discrepancy between the objective uncertainty $\pi$ and the agent's
initial perception of it. Formally, we denote by $\widetilde{\pi}$ the agent's
initial belief that the state is $1$ (or, expressed in odds ratio,
$\widetilde{\rho}$) and assume a stochastic relationship between
$\widetilde{\rho}$ and $\rho$:
\[
\widetilde{\rho}=\eta\rho
\]
where $\eta$ is a positive random variable.\footnote{In simulations to come,
we assume that $\log\eta$ is normally distributed.} When $\eta$ is
concentrated on $1$, the agent has correct priors.

For any belief $\widehat{\pi}$ about state 1 that the agent might hold upon
taking a decision, we assume that the agent chooses the welfare maximizing
action given this belief, i.e., chooses action $1$ when $\widehat{\pi
}(1-\gamma)>(1-\widehat{\pi})\gamma$, or equivalently, denoting $\widehat{\rho
}\equiv\widehat{\pi}/(1-\widehat{\pi})$ the odds ratio, when
\begin{equation}
\widehat{\rho}>\Gamma\label{EqD}%
\end{equation}
Thus, in the absence of any signals to be processed, the agent chooses action
1 when $\widetilde{\rho}>\Gamma$, hence on average across realizations of
$\widetilde{\rho}$, he obtains:
\[
\underline{W}\equiv(1-\gamma)\pi\Pr(\widetilde{\rho}>\Gamma)+\gamma(1-\pi
)\Pr(\widetilde{\rho}<\Gamma)
\]
In case the agent has correct priors, he achieves an expected welfare equal
to
\[
\underline{W}_{0}\equiv\max(\pi(1-\gamma),(1-\pi)\gamma)\geq\underline{W}%
\]

\subsection{Signals, evidence and strength of evidence}

For each problem faced, the agent receives, prior to making a decision, a
sequence of signals imperfectly correlated with $\theta$ that she may use to
form a posterior belief. The sequence is denoted $X$, assumed to be
arbitrarily long, and conditional on the true state $\theta$, each signal
$x\in X$ is drawn independently from the same distribution with density
$f(\cdot\mid\theta)$, assumed to be strictly positive and smooth on its
support $[0,1]$. In addition, the odd ratio
\[
L(x)\equiv f(x\mid1)/f(x\mid2)
\]
is assumed to be strictly increasing in $x$. The distributions $\{f(\cdot
\mid\theta)\}_{\theta}$ are \textit{problem specific} and we think of them as
objective characteristics of the problem faced.

When signal $x$\ arises, there is a state $\overline{\theta}(x)\in
\{1,2\}$\ that has highest likelihood (or that best fits $x$),
namely:\textbf{\ }%
\[
\overline{\theta}(x)=\arg\max_{\theta\in\{1,2\}}f(x\mid\theta).
\]
We say that signal $x$ is \textit{evidence} for state $\overline{\theta}%
(x)$.\footnote{Given our assumption on $L$, there is a unique uninformative
signal $x_{0}$ (i.e., $L(x_{0})=1$): all signals above $x_{0}$ provide
evidence for $\theta=1$, and all signals below $x_{0}$ provide evidence for
$\theta=2$.} To measure the \textit{strength of the evidence}, we define%
\[
l(x)=\frac{f(x\mid\theta=\overline{\theta}(x))}{f(x\mid\theta\neq
\overline{\theta}(x))}=\max(L(x),1/L(x)).
\]
Any signal $x$ thus has an ``objective" (informational) characteristics
$h\equiv(\overline{\theta},l).$ We shall denote by $H$ the sequence of
characteristics associated with the sequence $X$. Figure \ref{fig0z} provides
an illustration: \begin{figure}[h]
\centering
\begin{minipage}{0.45\textwidth}
\centering
\includegraphics[scale=0.45]{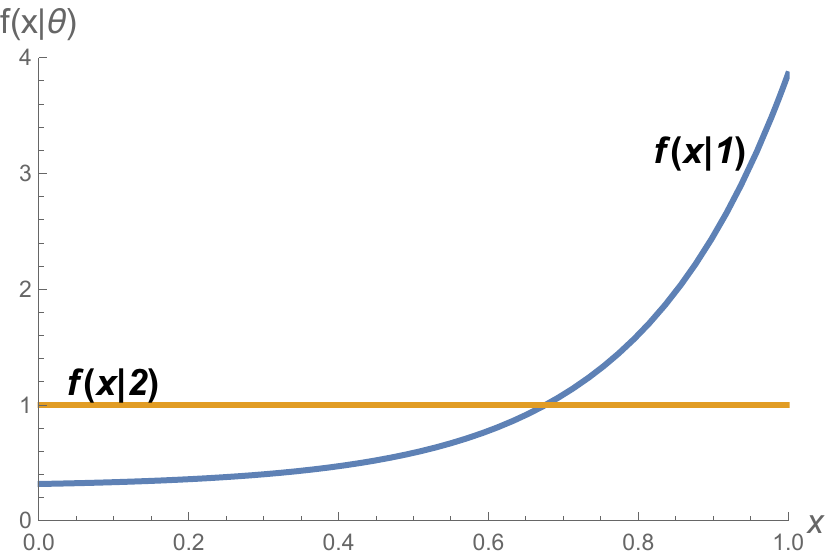}
\end{minipage}\hfill\begin{minipage}{0.45\textwidth}
\centering
	 \includegraphics[scale=0.45]{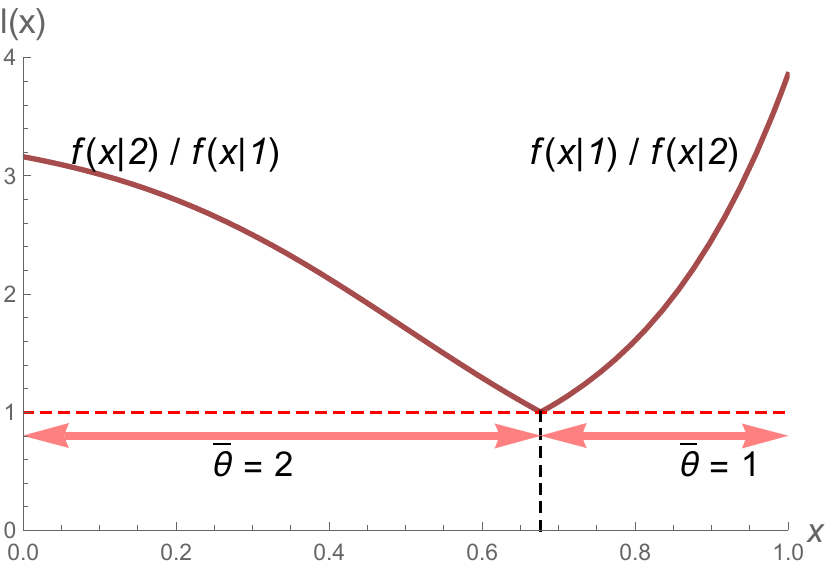}
\end{minipage}
\caption{From distributions to strength of evidence}%
\label{fig0z}%
\end{figure}

\subsection{Censoring and coarse processing.}

One aspect of our analysis will be the possibility that a signal does not get
to the agent's attention, or that it is simply not processed, for example
because its \textit{strength} is too weak, i.e., not informative enough.
Another aspect will be that even when a signal is processed, its informative
value is difficult to assess, noisy or coarse. Here we assume that
\textit{perceptions are coarse}:\footnote{In the discussion Section, we
briefly discuss cases where the agent's perception of $l$ is noisy.} the
signal is either \textit{not processed} ($\widetilde{\theta}=0$), or when
processed, it is either perceived as \textit{evidence in favor of} $\theta=1$
(in which case $\widetilde{\theta}\equiv1$) or as \textit{evidence in favor of
}$\theta=2$ (in which case $\widetilde{\theta}\equiv2$). So $\widetilde{\theta
}\in\{0,1,2\}$.

\begin{figure}[h]
\centering
\includegraphics[scale=0.6]{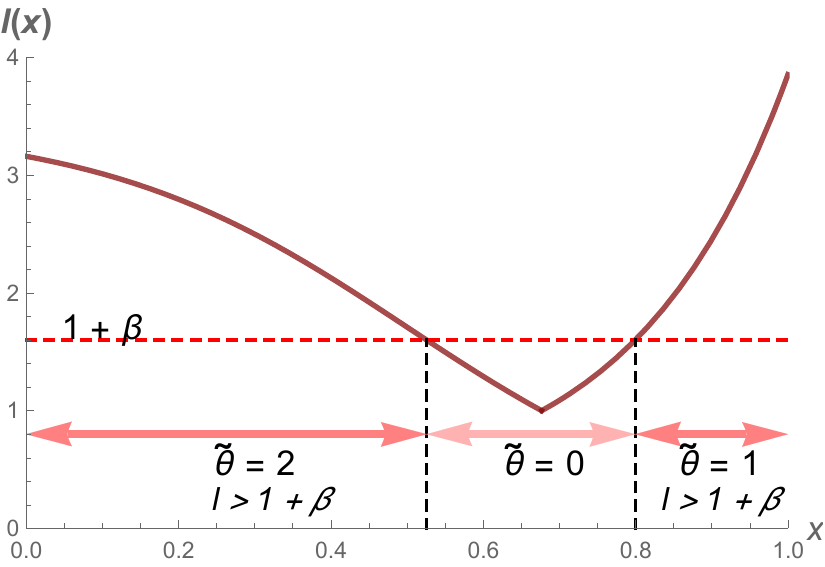}\caption{Censoring and coarse
processing}%
\label{f0zz}%
\end{figure}

Formally, we define the threshold strength $1+\beta$ above which the signal
gets to the agent's attention,\footnote{We shall later endogenize the
incentives to censor weak evidence.} and we denote by $\widetilde{X}_{\beta}$
the subsequence of signal actually processed:%
\[
\widetilde{X}_{\beta}=\{x\in X,l(x)\geq1+\beta\}
\]
where $\beta\geq0$ characterizes the degree to which weak signals are censored
or go unnoticed. We further assume that for uncensored signals, the direction
of evidence is correctly perceived, i.e. $\widetilde{\theta}=\overline{\theta
}$, but this is not central to our analysis. We shall denote by $\widetilde{H}%
$ the sequence of perceptions, which consists of a sequence of realizations of
$\widetilde{\theta}\in\{0,1,2\}$. Figure~\ref{f0zz} explains which perception
is generated for each $x$, for a given distribution $f$.

We also assume the agent takes a decision after $N$ signals have been
processed, and unless otherwise mentioned (i.e., in Section
\ref{SectionFewerSignals}), we consider the limit case where $N$ is
arbitrarily large.

\subsection{Coarse mental system and belief formation}

We consider agents attempting to form beliefs based on the sequence of
perceptions $\widetilde{H}$: (i) how should the agent aggregate the coarse
perceptions $\widetilde{\theta}$? (ii) Given this aggregation, what posterior
belief should the agent hold?

To answer (i) we posit the\textit{\ simple mental system} described in
Introduction. The agent starts at $s=0$, moving one step up (if possible and)
if $\widetilde{\theta}=1$, moving one step down (if possible and) if
$\widetilde{\theta}=2$, where $s\in S\equiv\{-K,..,0,...K\}$.

To answer (ii), we assume that when in state $s$ prior to decision making, the
agent uses a\textit{\ simple belief-formation strategy}:
\[
\sigma^{d}(\widetilde{\rho},s)=\widetilde{\rho}d^{s}%
\]
for some $d\geq1$. When $d>1$, a positive (negative) mental state thus moves
the agent away from her prior, towards believing that $\theta=1$ ($\theta=2$).
The parameter $d$ captures the degree to which the agent's mental system
influences beliefs. When $d=1$, the agent keeps her prior and effectively
ignores all signals received.

For the sake of exposition, we also introduce a more \textit{sophisticated
updating rule} in which the agent updates his initial belief $\widetilde{\rho
}$ according to
\[
\sigma^{d,\Lambda}(\widetilde{\rho},s)=\widetilde{\rho}\Lambda d^{s}.
\]
With rules of this kind, the agent has two instruments: the degree $d$ to
which her mental system influences beliefs, and the degree to which she biases
her decision: if her mental system tends to generate higher mental states on
average, she will have an incentive to choose $\Lambda$ below 1.

\subsection{Belief formation and welfare.}

We are interested in the welfare performance of censoring and belief-formation
strategies. For any fixed $f$, sequence $X$ and number $N$ of signals
processed, the censoring level $\beta$ determines a posterior mental state
$\widehat{s}$, hence, for a given prior realization $\widetilde{\rho}$ and
parameter $d$, a posterior belief $\widehat{\rho}=\widetilde{\rho
}d^{\widehat{s}}$. Using this posterior belief, the agent chooses action 1
when $\widehat{\rho}>\Gamma$ (see (\ref{EqD})), so taking expectations over
the realizations $X$ and $\widetilde{\rho}$ and computing the large $N$ limit,
the expected welfare is given by:%
\begin{equation}
W_{f}(\beta,d)=(1-\gamma)\pi\Pr\nolimits_{f,\beta,d}(\widehat{\rho}%
>\Gamma)+\gamma(1-\pi)\Pr\nolimits_{f,\beta,d}(\widehat{\rho}<\Gamma)
\label{Welfare}%
\end{equation}
We shall be interested in the performance of $\beta$ and $d$ on average over
the possible realizations of $f$, that is:%
\[
W(\beta,d)=E_{f}W_{f}(\beta,d)
\]

\subsection{The mental state dynamics.}

To conclude this Section, we explain how the data generating process $f$ and
the censoring parameter $\beta$ affect the mental state dynamics. We also
recall what we mean by regular and irregular problems.

\begin{figure}[h]
\centering
\begin{minipage}{0.45\textwidth}
\centering
\includegraphics[scale=0.47]{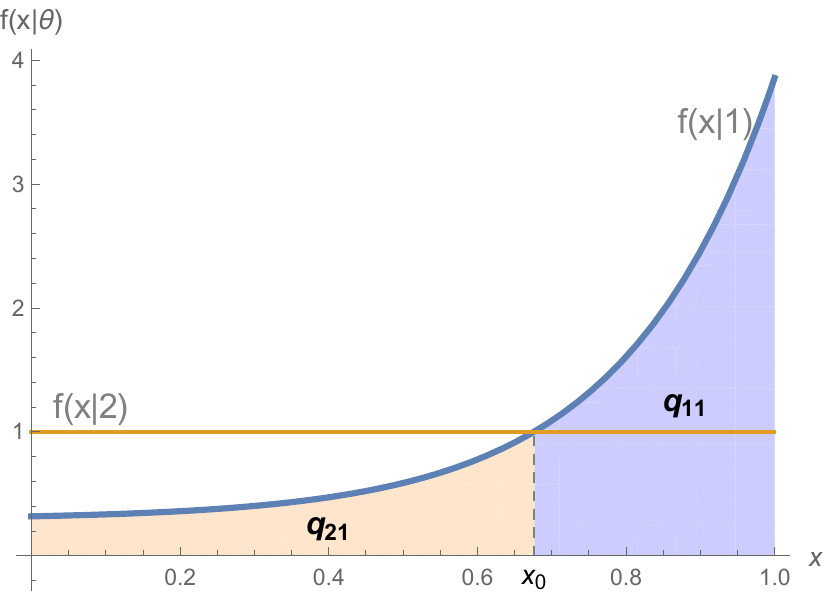}
\subcaption{$\theta=1$, $\beta=0$}
\end{minipage}\hfill\begin{minipage}{0.45\textwidth}
\centering
	 \includegraphics[scale=0.47]{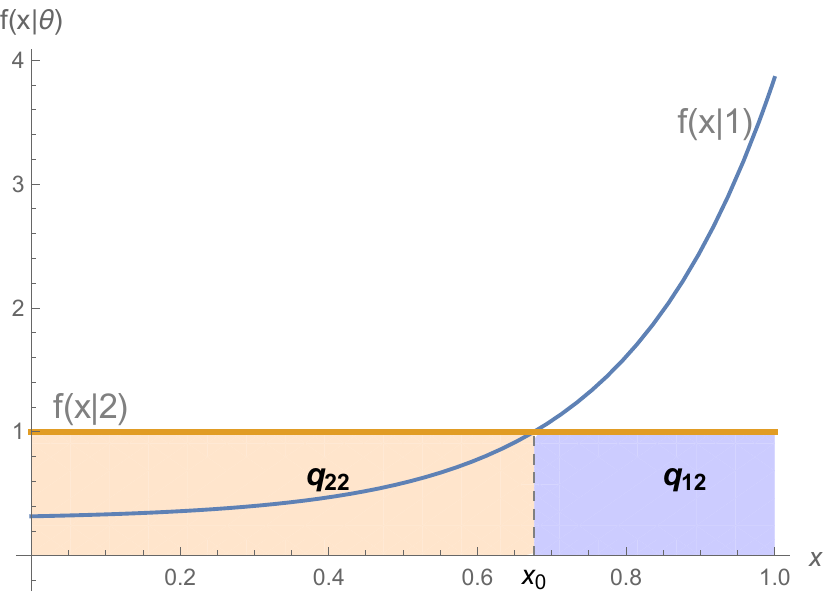}
\subcaption{$\theta=2$, $\beta=0$}
\end{minipage}
\par
\begin{minipage}{0.45\textwidth}
\centering
\includegraphics[scale=0.47]{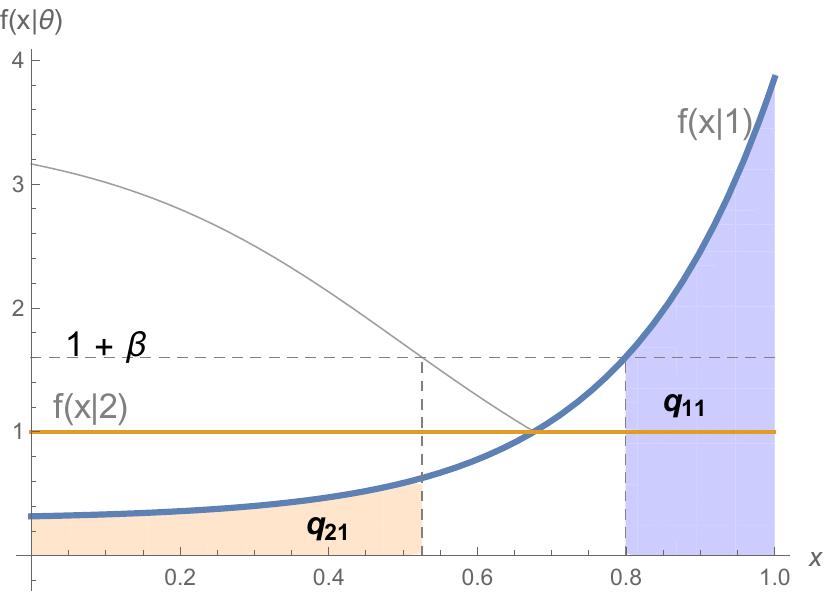}
\subcaption{$\theta=1$, $\beta=0.6$}
\end{minipage}\hfill\begin{minipage}{0.45\textwidth}
\centering
	 \includegraphics[scale=0.47]{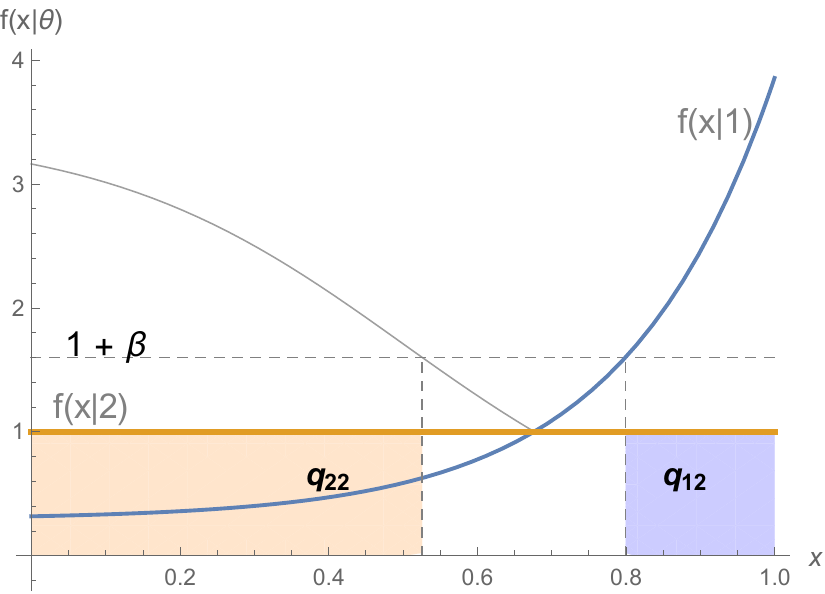}
\subcaption{$\theta=2$, $\beta=0.6$}
\end{minipage}
\caption{Transition probabilities}%
\label{fig1}%
\end{figure}

For given $f$ and $\beta$, the mental-state dynamic is entirely driven by the
vector of transition probabilities $q=(q_{\widetilde{\theta}\theta
})_{\widetilde{\theta},\theta}$ where%
\[
q_{\widetilde{\theta}\theta}\equiv\Pr(\widetilde{\theta}\mid\theta,f,\beta)
\]
is defined for $\theta\in\{1,2\}$ and $\widetilde{\theta}\in\{0,1,2\}$. Figure
\ref{fig1} below provides a graphic representation of these probabilities for
a given $f$, with $\beta=0$ (no censoring) or $\beta=0.6$. In each figure, the
blue area corresponds to the probability of moving upward ($\widetilde{\theta
}=1$). This probability depends on the underlying state $\theta$ (as this
defines the relevant distribution over signals $f(.|\theta)$) and the level of
censoring $\beta$.\footnote{The figures assume that conditional on processing,
attributions are correct $\widetilde{\theta}=\overline{\theta}$}

At the large $N$ limit, only the events where $\widetilde{\theta}\in\{1,2\}$
matter so we define the conditional probabilities
\[
p_{\widetilde{\theta}\theta}\equiv\frac{q_{\widetilde{\theta}\theta}%
}{q_{1\theta}+q_{2\theta}}\text{, for }\theta\in\{1,2\}\text{ and
}\widetilde{\theta}\in\{1,2\}.
\]
Since $p_{1\theta}+p_{2\theta}=1$, we can summarize the mental state dynamics
by the vector $p=(p_{11},p_{22})$. Furthermore, we shall distinguish between
regular and irregular problems:

\textbf{Definition}: \textit{We say that a problem is regular if }$p_{11}%
>1/2$\textit{ and }$p_{22}>1/2$\textit{. It is irregular otherwise.}

Finally, note that while the censoring level $\beta$ affects $p$, it does not
directly affect welfare: given $p$, welfare is only determined by the
belief-formation strategy $\sigma$, so for notational convenience, we shall
also write $W_{p}(\sigma)$ for the welfare induced by $\sigma$ when the mental
state dynamics is summarized by $p$.\footnote{That is, $W_{f}(\beta,d)\equiv
W_{p}(\sigma^{d})$ where $p$ is the vector of transitions induced by $f$ and
$\beta$.}


\section{Optimal belief formation}

As a benchmark case, we start by exploring optimal belief formation when all
belief formation rules are allowed, priors are correct ($\widetilde{\rho}%
=\rho$) and belief formation can be tuned to each specific $p$. This case
corresponds to the classic Bayesian case, under the constraints imposed by
censoring and mental processing. We next move to the case where belief
formation cannot be tuned to $p$.

\subsection{p-optimal strategies and the Bayesian case}

For any given $p$ and $\theta\in\{1,2\}$, denote by $\theta^{\prime}%
\in\{1,2\}\ $the alternative state, and define%
\[
d_{\theta}^{p}\equiv\frac{p_{\theta\theta}}{p_{\theta\theta^{\prime}}}\text{
and }r_{\theta}^{p}\equiv\frac{p_{1\theta}}{1-p_{1\theta}}%
\]
The ratio $d_{\widetilde{\theta}}^{p}$ characterizes, \textit{from a Bayesian
perspective}, the informativeness of perception $\widetilde{\theta}$, i.e. the
degree to which it supports $\theta=\widetilde{\theta}$, as opposed to the
alternative state $\theta^{\prime}\neq\theta$. The ratio $r_{\theta}^{p}$
characterises the degree to which the mental system points towards state $1$
(hence higher mental states) under state $\theta$. We further
define:\footnote{Note that since $p_{\theta\theta}=1-p_{\theta^{\prime}\theta
}$, we have $d_{p}=\frac{p_{11}}{p_{21}}\frac{p_{22}}{p_{12}}=\frac{p_{11}%
}{p_{21}}/\frac{p_{12}}{p_{22}}=r_{1}^{p}/r_{2}^{p}$, so both $d_{p}$ and
$\Lambda_{p}$ can be expressed as functions of $r_{\theta}^{p}$. We choose
this formulation to relate $d_{p}$ to the informativeness of each perception
$\widetilde{\theta}$ (for a Bayesian).}
\[
d_{p}=d_{1}^{p}d_{2}^{p}\text{ and }\Lambda_{p}=\sum_{s\in S}(r_{2}^{p}%
)^{s}/\sum_{s\in S}(r_{1}^{p})^{s}%
\]
We have:

\textbf{Proposition 1}: \textit{Assume the agent has correct priors. For any
fixed }$p,$\textit{\ the strategy }$\sigma_{p}^{\ast}\equiv\sigma
^{d_{p},\Lambda_{p}}$\textit{\ achieves maximum welfare across all possible
belief-formation rules }$\sigma$\textit{.}

Intuitively, the rule $\sigma^{d_{p},\Lambda_{p}}$ corresponds to Bayesian
updating given the constraint imposed by coarse perceptions and simple mental
processing. $\Lambda_{p}$ characterizes the informational content of being in
state $0$ in the long-run and $d_{p}$ characterizes the informativeness of
being in a higher mental state (by one step). When $\Lambda_{p}<1$, this means
that, for a Bayesian, being in mental state $0$ is evidence for $\theta=2$.
This happens when the mental system is \textquotedblleft%
\textit{unbalanced\textquotedblright}, i.e., leaning towards higher mental
states on average, and a Bayesian (who understands the process that generates
perceptions and mental states) corrects for this asymmetry. The proof is in
the Appendix.

\begin{figure}[h]
\centering
\begin{minipage}{0.4\textwidth}
\centering
\includegraphics[scale=0.4]{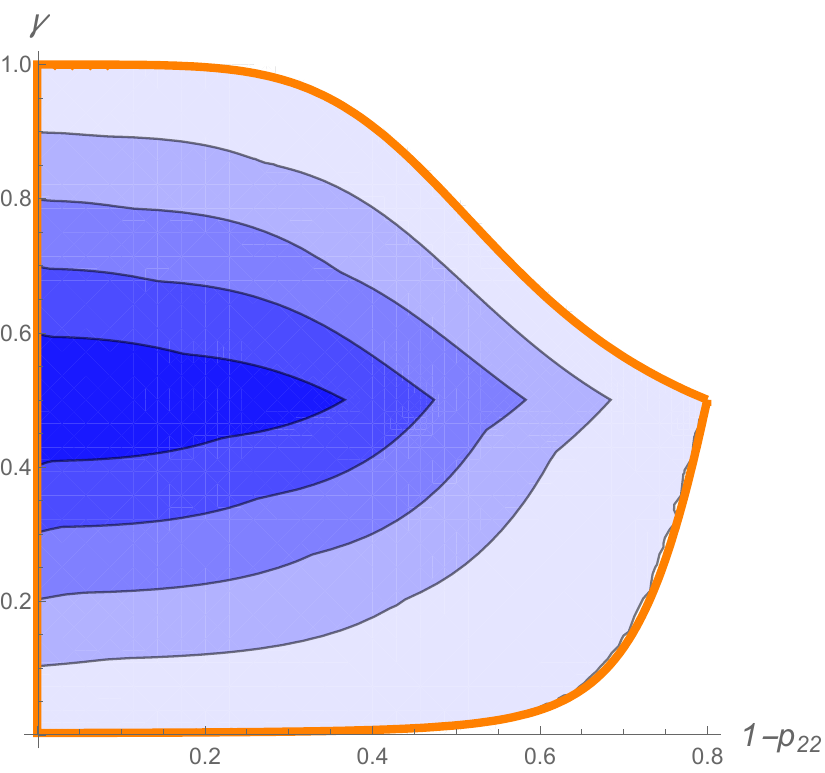}
\subcaption{Fixed $p_{11}=0.8$}
\end{minipage}\hfill\begin{minipage}{0.4\textwidth}
\centering
	 \includegraphics[scale=0.4]{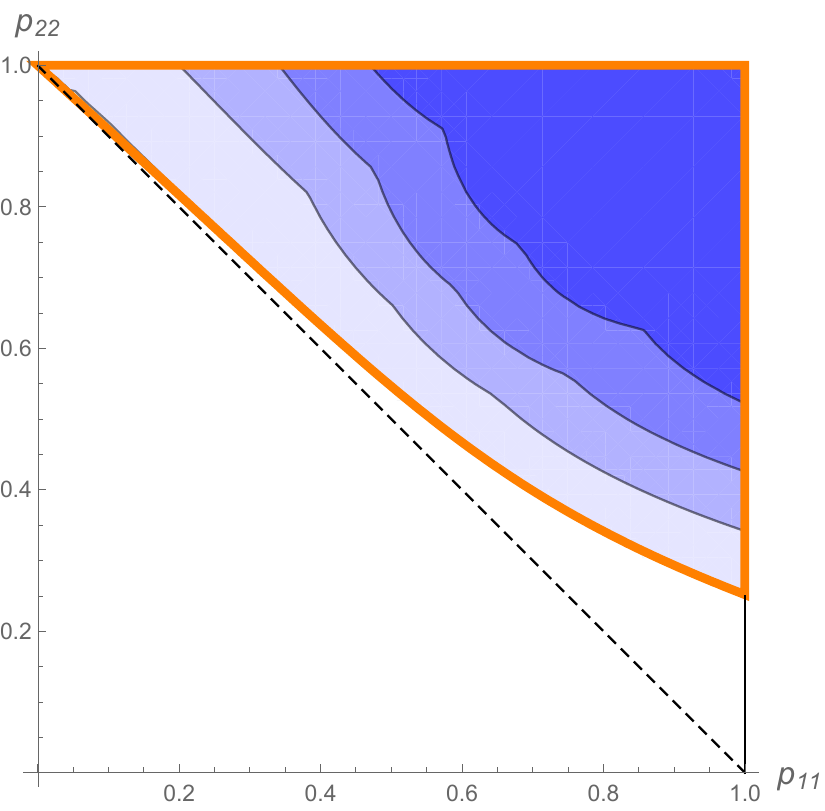}
\subcaption{Fixed $\gamma=0.6$}
\end{minipage}
\hfill\begin{minipage}{0.1\textwidth}
\centering
	 \includegraphics[scale=0.4]{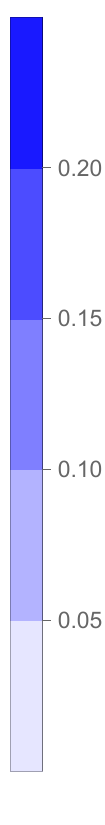}
\end{minipage}
\caption{Welfare gains}%
\label{fig2}%
\end{figure}

To assess the magnitude of the welfare gains, we assume $K=2$ (five mental
states) and compute numerically the welfare gains
\[
\Delta(p)\equiv W_{p}(\sigma_{p}^{\ast})-\underline{W}%
\]
associated with $\sigma_{p}^{\ast}$ compared to only relying the prior. With
correct priors, by Proposition 1, the agent cannot be worse off using
$\sigma_{p}^{\ast}$ so $\Delta(p)$ is non negative, and whenever $\sigma
_{p}^{\ast}$ induces an action that differs from what the prior suggests,
welfare must increase strictly.
Fixing $\pi=1/2$, Figure \ref{fig2}a and \ref{fig2}b report the domain of
strict welfare gains when $p_{11}=0.8$ (a) and when $\gamma=0.6$ (b), as well
as the magnitude of these gains.

The orange line defines the boundary of the domain for which
$\Delta(p)$ is strictly positive. Outside this domain, the decision maker
plays the same action irrespective of her mental state: the mental system is
not informative enough to tilt the decision away from what the prior suggests.
This happens when $1-p_{22}$ is too close to $p_{11}$, as the informativeness
of the mental system is then small (i.e., $d_{p}$ is close to 1) or when
$\gamma$ lies away from $1/2$, as substantial evidence is then required to
override the prior.

Figure~\ref{fig2} corroborates the standard insight that processing signals
correlated with the underlying state cannot hurt welfare. With noisy priors,
$\sigma_{p}^{\ast}$ is not the welfare optimizing rule. Nevertheless,
Figure~\ref{fig3} below shows that comparable gains obtain in that case as
well. \begin{figure}[h]
\centering
\begin{minipage}{0.4\textwidth}
\centering
\includegraphics[scale=0.4]{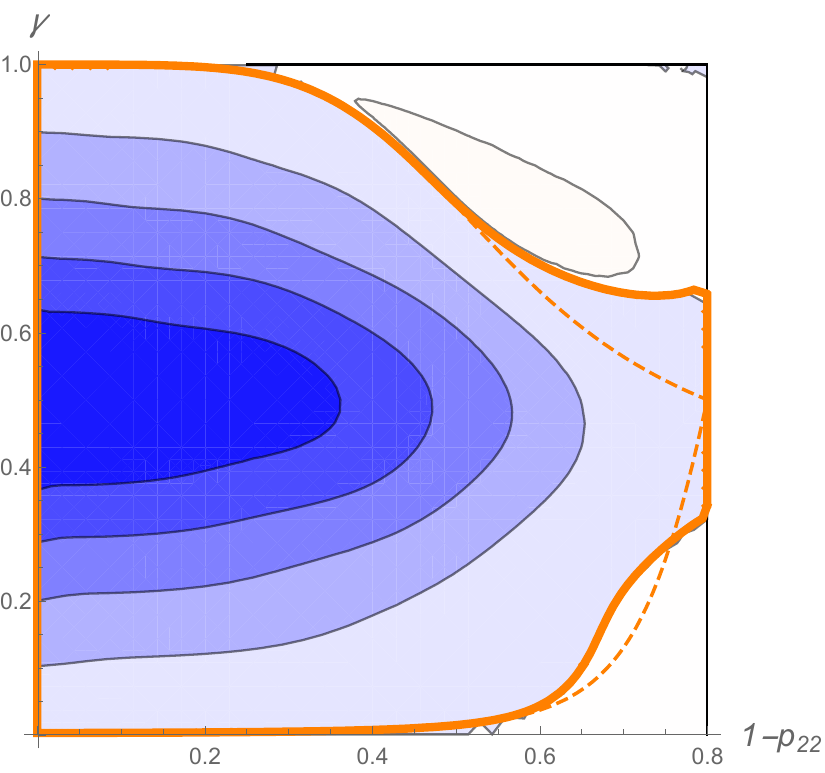}
\subcaption{Fixed $p_{11}=0.8$}
\end{minipage}\hfill\begin{minipage}{0.4\textwidth}
\centering
	 \includegraphics[scale=0.4]{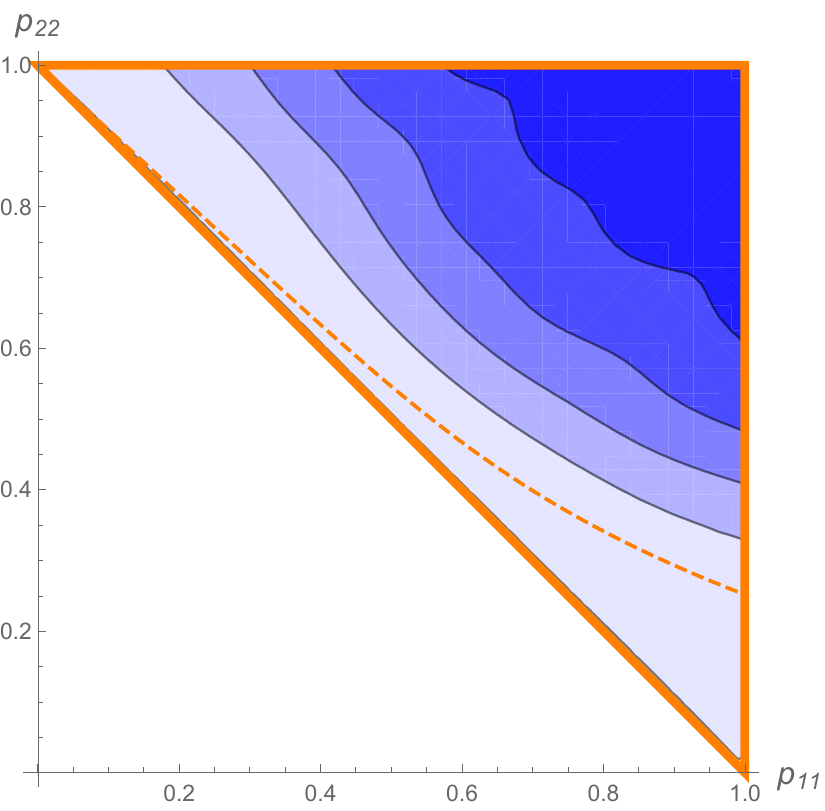}
\subcaption{Fixed $\gamma=0.6$}
\end{minipage}
\hfill\begin{minipage}{0.1\textwidth}
\centering
	 \includegraphics[scale=0.4]{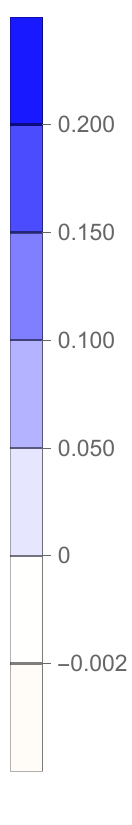}
\end{minipage}
\caption{Welfare gains under noisy priors}%
\label{fig3}%
\end{figure}

Intuitively, when priors are noisy, there are two effects at work: (i) relying
on priors is a worse option than before ($\underline{W}<\underline{W}_{0})$;
(ii) for a fixed $(\pi,\gamma)$, the change in belief required to switch
decision is modified. Observation (i) implies an expansion (in most
directions) of the set of parameters for which $\sigma_{p}^{\ast}$ helps
compared to relying on priors only, as well as comparable welfare gains for
most parameters. Observation (ii) implies that (for a small range of
parameters), the agent may be (slightly) worse off using the mental system
than his (noisy) prior. For these parameters, the agent has the illusion that
the mental system is powerful enough to override the prior, while he would be
better off ignoring the mental system.\footnote{When priors are noisy,
$\sigma_{p}^{\ast}$ is not the welfare optimizing rule. This is why
$\Delta(p)$ can be negative.}

\subsection{When $d$ cannot be adjusted to $p$}

We now explore the case where priors are noisy and the updating strategy
cannot be tuned to $p$. We make two observations. First, for any given $d>0$,
the strategy $\sigma^{d}$ works well for many problems. Second, for some
problems, the agent would be better off ignoring signals and only trusting her prior.

Formally, consider the welfare gain $\Delta^{d}(p)\equiv W_{p}(\sigma
^{d})-\underline{W}$ associated with \textit{using a given strategy}
$\sigma^{d}$. We observe that for a range of parameters $p$, $\Delta^{d}$ is
positive irrespective of $d$. Formally, let
\begin{equation}
B=\{p,d_{p}>\max(\frac{\Gamma}{\rho\Lambda_{p}},\frac{\rho\Lambda_{p}}{\Gamma
})\label{eqA}%
\end{equation}
\textbf{Proposition 2}: \textit{For any }$p\in B$\textit{, }$\Delta^{d}%
(p)>0$\textit{ for all }$d>0.\smallskip$

The set $B$ corresponds to cases where the flow of evidence remains somewhat
balanced $(\Lambda_{p}$ not too far from $1$) and the signals are sufficiently
informative $(d_{p}$ is large enough).\footnote{Conversely, the condition
$p\in B$ $\ $is more difficult to satisfy when stakes or priors are favoring
too much a given alternative, or when the direction of evidence points in the
same direction on average, independently of the underlying state.}
Intuitively, when $p\in B$, perceptions $\widetilde{\theta}\in\{1,2\}$ are
sufficiently strong evidence (in a Bayesian sense) in favor of state
$\theta=\widetilde{\theta}$ to ensure that when many signals have been
processed, any positive (respectively negative) mental state is correlated
with $\theta=1$ (respectively $\theta=2$). As a consequence, for any monotonic
belief strategy $\sigma$, welfare gains are positive \textit{for any non-zero
mental state reached.}

Of course, $\Delta^{d}$ can be positive under milder conditions, as welfare
gains need not be positive for each non-zero mental state possibly reached:
mostly matters states that are \textit{more likely} to be reached, so welfare
may increase over a range of $p$ larger than $B$ (see Figure \ref{fig4}
below). In contrast to the Bayesian case however, there is now a significant
range of parameters for which $\sigma^{d}$ hurts welfare. The reason is that
for some $p$, the mental system may generate evidence towards
$\widetilde{\theta}=1$ irrespective of the underlying state, and $\sigma^{d}$
does not correct for that\footnote{This is unlike a Bayesian would set $d$ and
$\Lambda<1$ appropriately.}. Figure \ref{fig4}a and \ref{fig4}b below
illustrate these observations assuming $d=3$ and $p_{11}=0.8.$\footnote{As
before, we fix $\pi=1/2$, $p_{11}=0.8$, $K=5$, so problems are parameterized
by $(p_{22},\gamma)$. Priors are noisy with $Log\mu\sim\mathcal{N}(0,0.5)$.}

\begin{figure}[h]
\centering
\begin{minipage}{0.4\textwidth}
\centering
\includegraphics[scale=0.4]{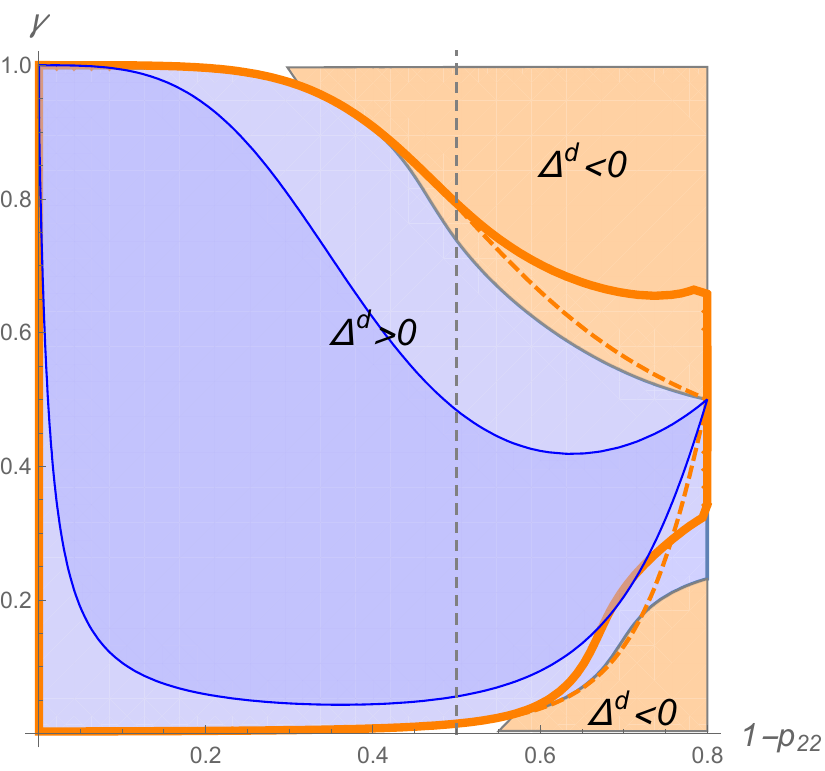}
\subcaption{Sign of $\Delta^d$}
\end{minipage}\hfill\begin{minipage}{0.4\textwidth}
\centering
	 \includegraphics[scale=0.4]{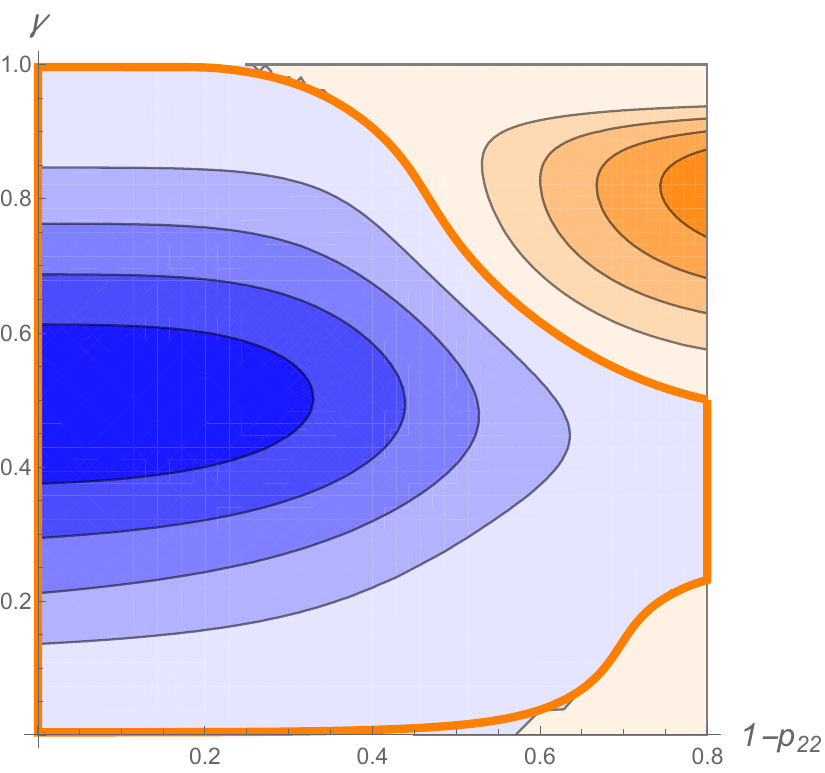}
\subcaption{Magnitude of $\Delta^d$}
\end{minipage}
\hfill\begin{minipage}{0.1\textwidth}
\centering
	 \includegraphics[scale=0.4]{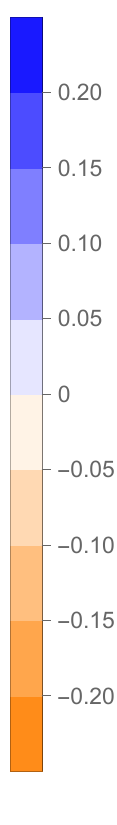}
\end{minipage}
\caption{Welfare gains untuned to $d.$}%
\label{fig4}%
\end{figure}The orange boundary recalls the domain defined in Fig \ref{fig3}a
for which the Bayesian strategy improves decision making. In the left figure,
the blue domain indicates the set of problems for which $\sigma^{d}$ improves
decision making, and the darker blue region defines domain $B$. In the rest of
the domain, using $\sigma^{d}$ decreases welfare, and the right figure
provides the magnitude of these changes: the magnitude of the losses can be as
large as that of the gains.

In the absence of optimization tuned to $p$, one expects that agents end up
with mistaken beliefs for some (irregular) problems: they will form a
posterior belief that bents towards one state of the world, mistakenly
thinking that their mental system permits to discriminate well between states
of the world, while their mental state primarily results from the fact that
evidence on average points towards the same direction irrespective of the
underlying state.

\section{Incentives to ignore weak evidence\label{SectionWeak}}

Ignoring weak evidence modifies the distribution of signals processed.
Starting from a situation where signals are not censored ($\beta=0$), we first
study the effect on transition probabilities $p$ between mental states. Then,
as a benchmark, we study the welfare consequence in the Bayesian case (where
belief formation can be tuned to $p$). Next we study the welfare consequences
for a fixed $d$.

\subsection{How censoring affects $p.$}

Let us first illustrate graphically how transition probabilities are affected
by censoring, for a small $\beta$. \begin{figure}[h]
\centering
\begin{minipage}{0.45\textwidth}
\centering
\includegraphics[scale=0.4]{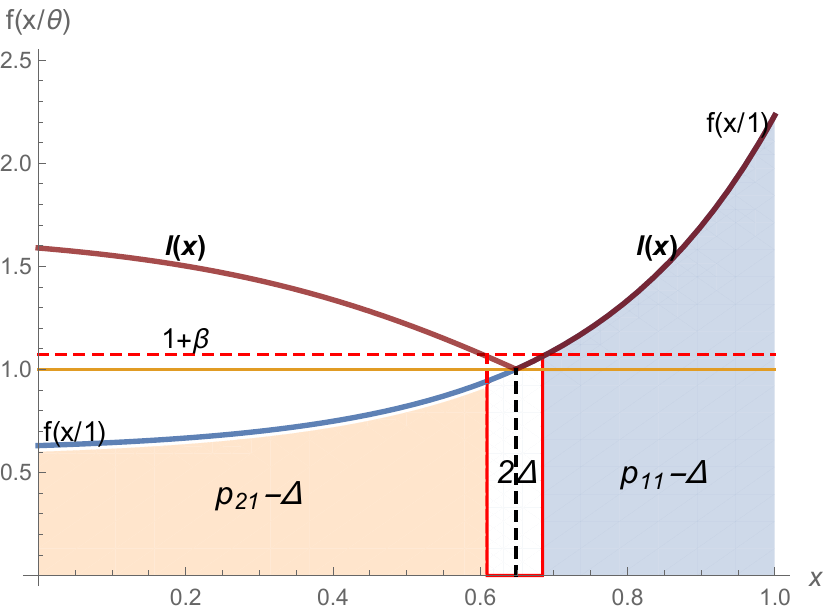}
\subcaption{Under $\theta=1$}
\end{minipage}\hfill\begin{minipage}{0.45\textwidth}
\centering
	 \includegraphics[scale=0.4]{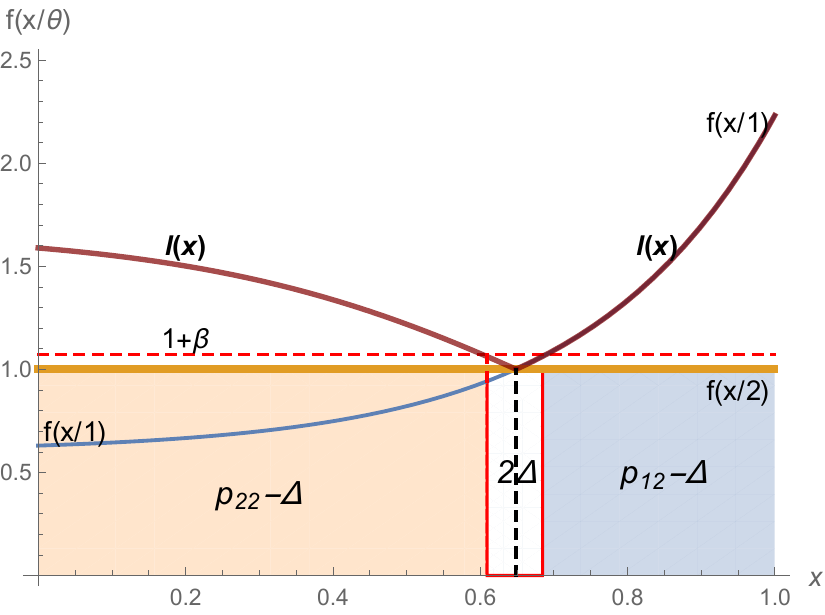}
\subcaption{Under $\theta=2$}
\end{minipage}
\caption{How censoring affects $p$}%
\label{fig5}%
\end{figure}
The chance that a signal is not processed is small, and denoted $2\Delta$ (see
Figure \ref{fig5}). One key observation is that when $\beta$ is small, the
weak evidence censored is equally likely to favor $\theta=1$ or $\theta=2$,
implying that both $p_{11}$ and $p_{21}$ are reduced by $\Delta$ (up to second
order terms).\footnote{This observation relies on our assumption that $L$ is
smooth and strictly increasing.} Said differently, processing weak evidence is
equivalent to adding state-independent noise to the mental system. This
observation implies (see Appendix):\smallskip

\textbf{Proposition 3.} \textit{At }$\beta=0$, (i) $\frac{\partial d_{p}%
}{\partial\beta}>0;$(ii) $\frac{\partial p_{kk}}{\partial\beta}\ $\textit{has
the same sign as} $p_{kk}-1/2$\textit{, and (iii) }$\frac{\partial\Lambda_{p}%
}{\partial\beta}>0$\textit{\ iff }$\Lambda_{p}>1$\smallskip

So when weak information is censored, $d_{p}$ rises, which implies that from a
Bayesian perspective, the spread in posterior beliefs is larger. But it also
implies that for most problems (i.e., unless $p_{11}=p_{22}$), $\Lambda_{p}$
lies further away from $1$, that is, the mental system is less balanced.

As we shall show, the consequence of the larger spread is that, in the
Bayesian case (where the agent can tune his strategy to $p$), censoring weak
evidence improves welfare for most values of $p$.\footnote{For most pairs, but
not for all, as we shall explain.} The consequence of the reduced balancedness
of the mental system is that when the agent follows a given strategy
$\sigma^{d}$ (which does not correct for this imbalance), censoring weak
evidence hurts welfare for some problems.

Nevertheless, we will show that $\sigma^{d}$ improves welfare for \textit{all}
``\textit{regular problems}", that is, problems for which%
\begin{equation}
p_{11}>1/2\text{ and }p_{22}>1/2 \tag{R}\label{R}%
\end{equation}
One may thus conclude that to the extent that regular problems are
preponderant, incentives to censor weak evidence are present even when the
agent cannot finely tune his belief-formation strategy to $p$.

\subsection{The Bayesian case\label{subsectionBayesianCaseWeakEvidence}}

\begin{figure}[h]
\centering
\begin{minipage}{0.45\textwidth}
\centering
\includegraphics[scale=0.45]{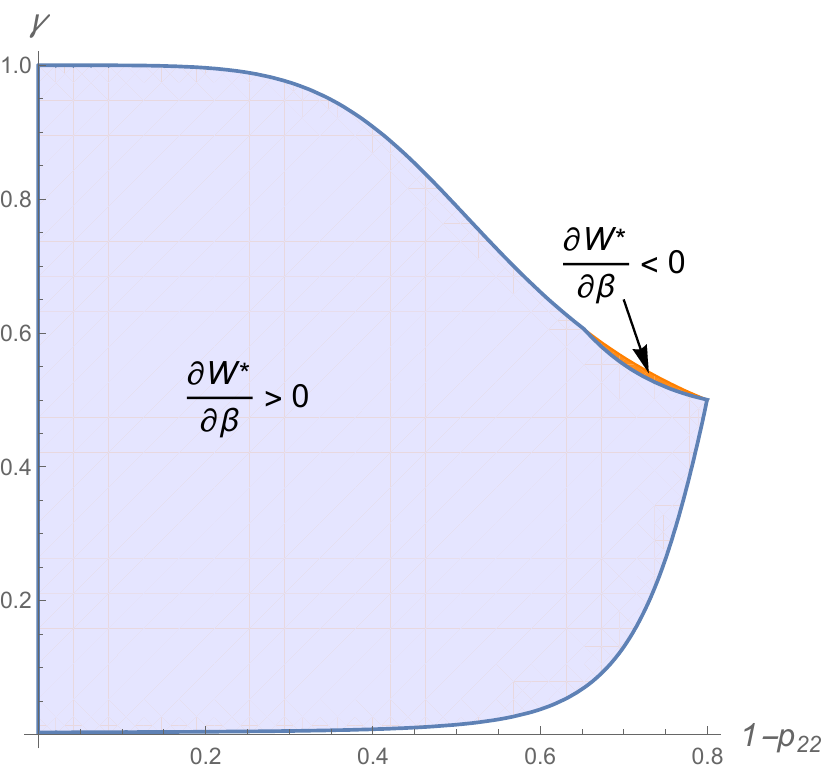}
\subcaption{case $p_{11}=0.8$}
\end{minipage}\hfill\begin{minipage}{0.45\textwidth}
\centering
	 \includegraphics[scale=0.45]{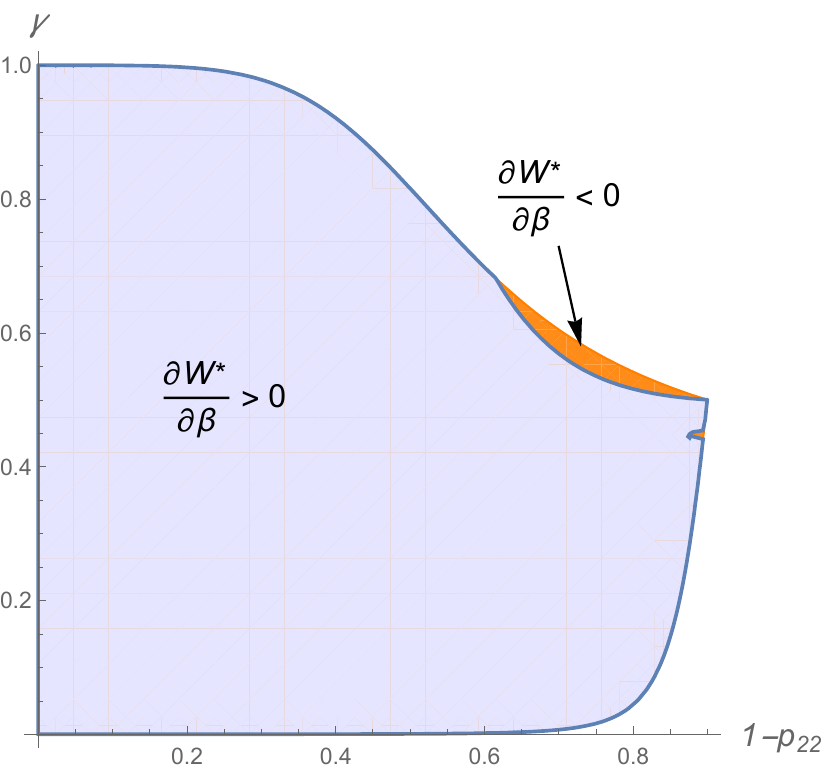}
\subcaption{case $p_{11}=0.9$}
\end{minipage}
\caption{Welfare gains from censoring: Bayesian case}%
\label{fig6}%
\end{figure}

When weak evidence is censored, the agent is relying on more informative
perceptions, so it would seem that, at least from a Bayesian perspective, this
always improves welfare. In particular, since $d_{p}$ rises, the spread in
posterior beliefs must increase, so the set of problem for which the mental
system helps would seem to increase as well. Figures \ref{fig6}a and
\ref{fig6}b plot the locus of problems for which censoring help (blue) and
hurts (orange) for two values of $p_{11}$. They illustrate that the intuition
above is correct, \textit{for most problems}. Interestingly however, we also have:

\textbf{Proposition 4.} \textit{In the Bayesian case, the set of problems and
priors for which censoring weak evidence hurts welfare is not empty.}

Intuitively, noise helps for example in problems where evidence is strongly
pointing towards the same direction, say $\widehat{\theta}=1$, so that in the
long run, the agent's mental state is generally close to the largest feasible
mental state $s=K$. In this case, adding noise to the mental system allows the
agent to better discriminate between the two underlying states, while
censoring evidence reinforces the concentration on the largest mental state
$K$ whether the state is $\theta=1$ or 2, reducing the informativeness of
being in $K$.

Formally, the proof consists in defining $\overline{\Lambda}_{p}=\Lambda
_{p}(d_{p})^{K}$ and showing that the set $D=\{p,\frac{\partial\overline
{\Lambda}_{p}}{\partial\beta}<0\}$ is non-empty. Then, although censoring
increases $d_{p}$, the largest possible shift in posterior beliefs decreases,
so for these marginal problems where $\Gamma/\rho$ is below but close to
$\overline{\Lambda}_{p}$, censoring makes the mental system useless.

\subsection{When $d$ cannot be tuned to $p$}

\begin{figure}[h]
\centering
\begin{minipage}{0.45\textwidth}
\centering
\includegraphics[scale=0.45]{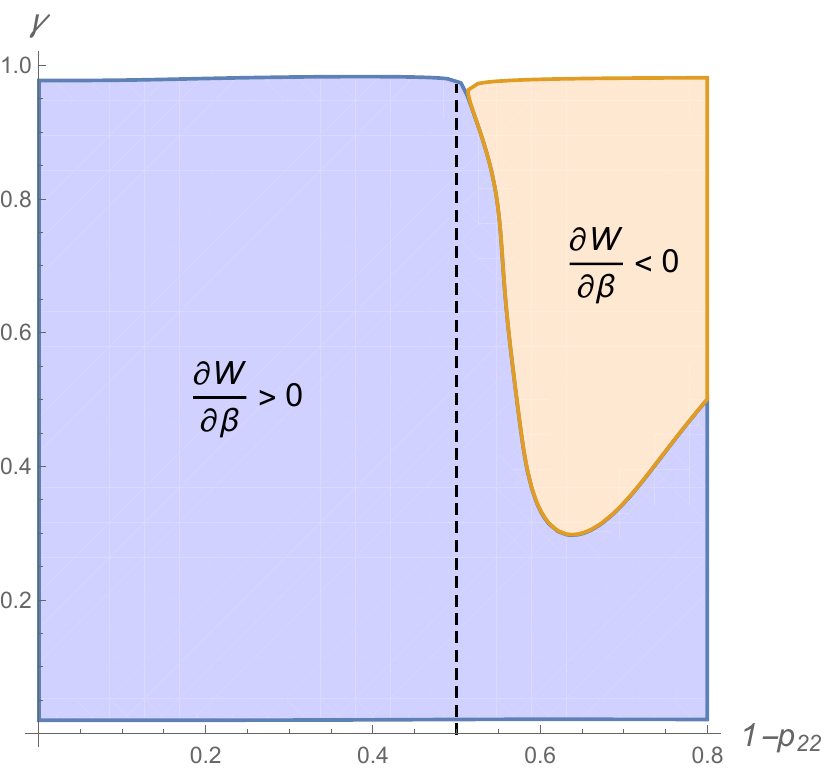}
\subcaption{fixed $p_{11}=0.8$}
\end{minipage}\hfill\begin{minipage}{0.45\textwidth}
\centering
	 \includegraphics[scale=0.45]{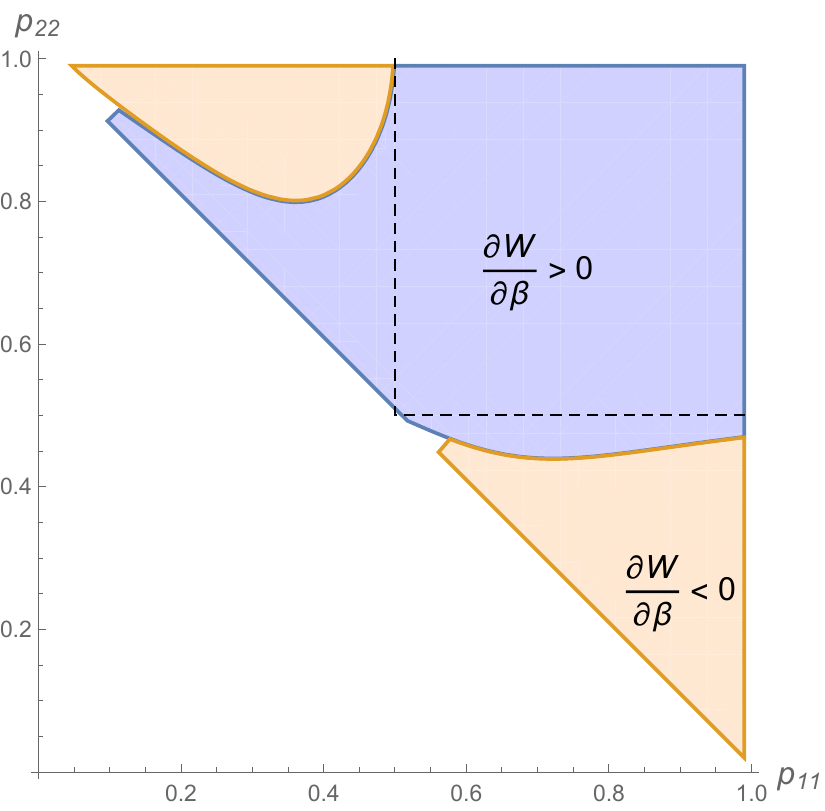}
\subcaption{fixed $\gamma=0.8$}
\end{minipage}
\caption{Welfare gains and losses from censoring (untuned case)}%
\label{fig7}%
\end{figure}

As we have just explained, ignoring weak evidence may further increase the
imbalance of the mental system. This implies that for simple belief-formation
strategies, which do not correct for this imbalance, and which are not tuned
to $p$, welfare may decrease. Figure \ref{fig7} confirms this and shows a
negative effect of censoring on welfare for a significant range of
problems.\footnote{On the left, we set $p_{11}=0.8$ and $\pi=1/2$, and examine
whether welfare increases (blue) or decreases (orange) depending on parameters
$p_{22}$ and $\gamma$. On the right, we fix $\gamma=0.8$, and examine
variations in the $(p_{11},p_{22})$ space.} Nevertheless, we show below that for \textit{all regular problems}
(both $p_{\theta\theta}$ above $1/2$)\footnote{The frontiers of regular
problems are indicated by the dashed line in the Figures. These problems lie
within the blue (welfare improving) region.}, and in spite of the increased
imbalance that censoring generates, welfare increases. \smallskip

\textbf{Proposition 5.} \textit{For any monotone belief formation strategy
}$\sigma$ \textit{and any regular problem, censoring weak evidence marginally
increases welfare.}\smallskip

Intuitively, the reason is that for these problems, ignoring weak evidence
always increases both $p_{11}$ and $p_{22}$, so it increases the correlation
between the underlying state $\theta=1$ (respectively $\theta=2$) and being in
a positive mental state (respectively negative mental state). The formal proof
is in the Appendix.

\section{The persistence of superstitions.}

\subsection{The role of asymmetries.}

Our hypothesis is that it is difficult for agents to adjust censoring and the
belief-formation rule to each problem that one faces, that is, to unobservable
characteristics of the data generating process. We think of these kinds of
adjustment as being more plausibly made on average across problems.

Given this hypothesis, the general message conveyed by the previous Sections
is that, to the extent that agents face a substantial fraction of regular
problems, agents have incentives to both censor weak evidence $(\beta>0$) and
to raise the power of their mental system ($d>1$).\footnote{Note that
censoring weak evidence actually increases the benefits of raising $d$ for
regular problems, because more informative signals are processed
}

The adverse consequence however is that for some problems, agents would be
better off not trusting their mental process and ignoring the updating that it
suggests. We illustrate below the type of problems for which this occurs.

Let us first observe the consequence of raising $\beta$ on $(p_{11},p_{22})$
for two different distributions $f(.|1)$. In the left figure, as $x$ departs
from the uninformative signal $x_{0}$, the strength of evidence $l(x)$ rises
in somewhat comparable ways (whether signals provide evidence in favor of
$\theta=1$ or $2$). As a result, both $p_{11}$ and $p_{22}$ rise. For the more
asymmetric distribution $f(.\mid1)$ considered on the right figure, this is
not the case, and a similar rise in $\beta$ sends $p_{22}$ to $0$: at
$\beta=1$, all signals processed are evidence in favor of $\theta=1$.

\begin{figure}[h]
\centering
\begin{minipage}{0.45\textwidth}
\centering
\includegraphics[scale=0.45]{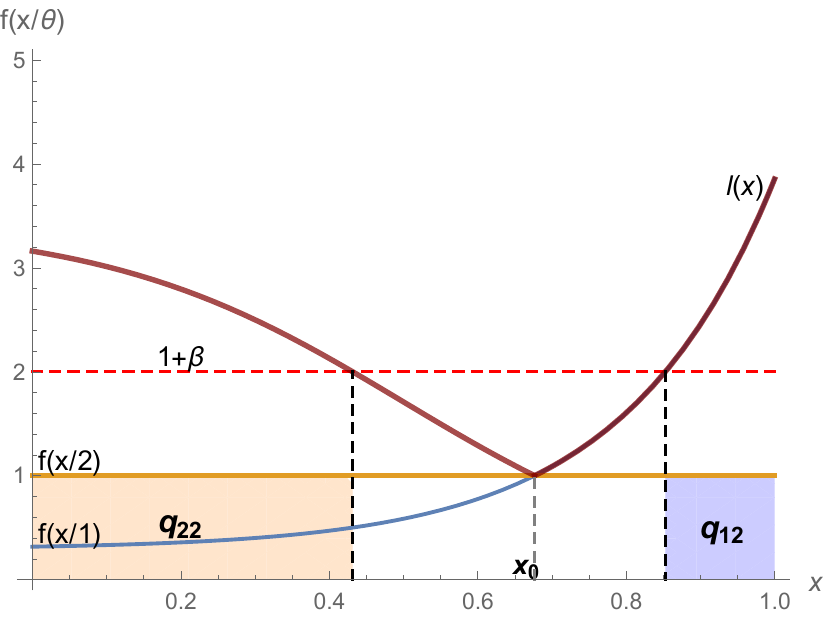}
\subcaption{$\theta=2$}
\end{minipage}\hfill\begin{minipage}{0.45\textwidth}
\centering
	 \includegraphics[scale=0.45]{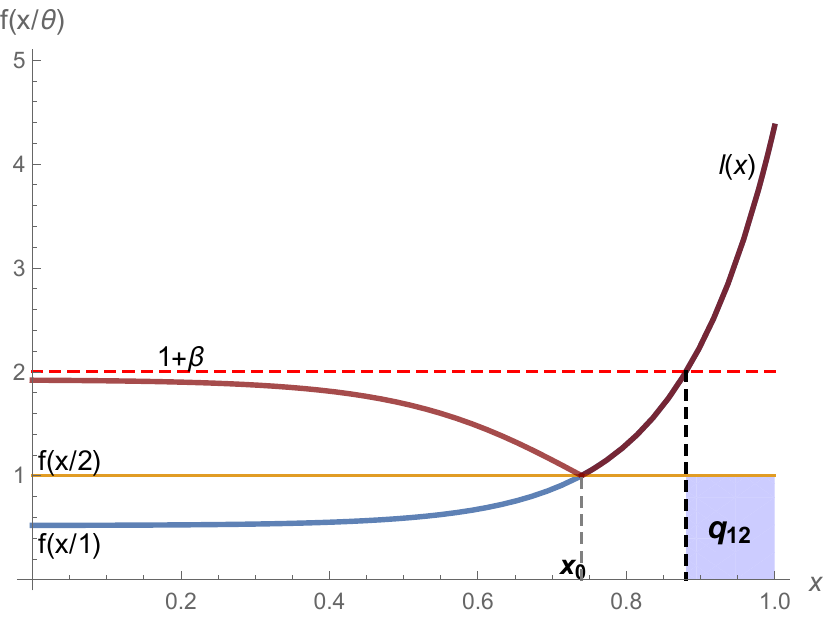}
\subcaption{$\theta=2$, asymmetric distributions}
\end{minipage}
\caption{Transition probabilities}%
\label{fig8}%
\end{figure}

\begin{figure}[h]
\centering \hfill\hfill\begin{minipage}{0.35\textwidth}
\includegraphics[scale=0.45]{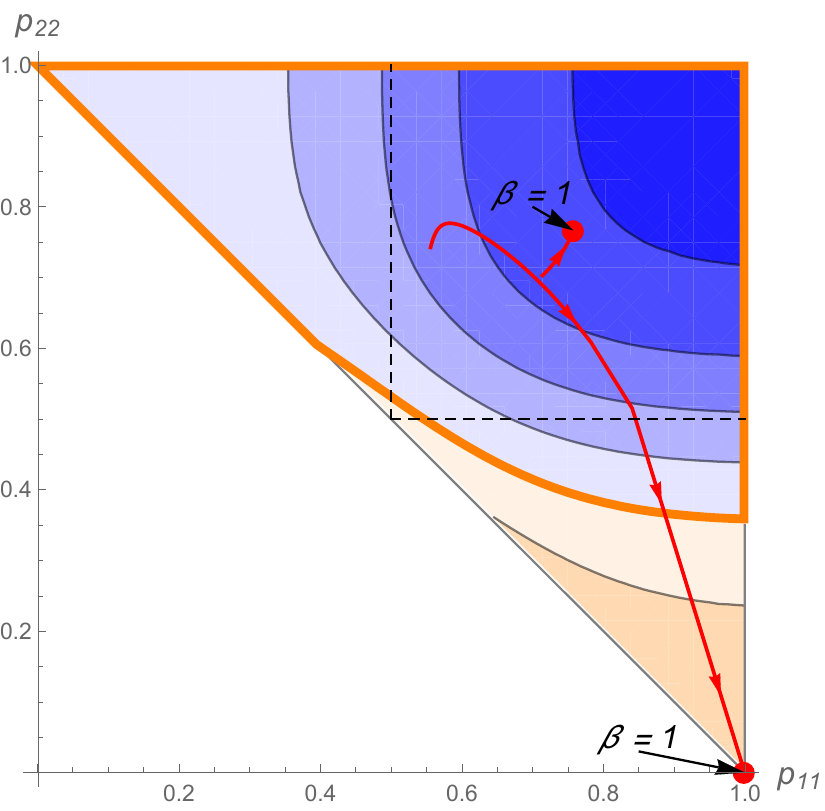}
\end{minipage}
\begin{minipage}{0.4\textwidth}
\centering
\includegraphics[scale=0.4]{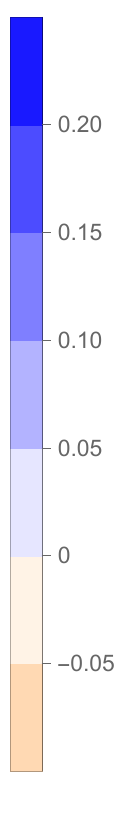}
\end{minipage}
\caption{When raising $\beta$ fosters biased beliefs}%
\label{fig9}%
\end{figure}

Next, for each distribution $f(.|1)$, we consider the effect of raising
$\beta$ from $0$ to $1$. Figure~\ref{fig9} indicates the path induced by this
rise in the $p$-space (i.e., the two red curves, one for each distribution).
Figure~\ref{fig9} also indicates welfare levels as a function of $p$ (for
$\gamma$ set to $0.6$).

Note that at $\beta=0$, marginally censoring weak evidence
generates a welfare gain for both problems. The consequence of more
significant censoring differs across problems. For the more \textquotedblleft
symmetric" distribution, welfare continues to increase at $\beta=1$. For the
more asymmetric distribution, $\beta=1$ sends welfare to the worst possible level.

\subsection{Examples.}

How does this relate to superstition, superstitious beliefs, or more
generally, folk beliefs? Our claim is that such beliefs typically arise for
problems of the asymmetric kind described above.

\paragraph{\textit{Lunar effects.}}

Consider an individual trying to discriminate between two states of the world.
Under state 1, full moon generates on average a 20\% increase in the number of
deliveries, while under state 2, there is no effect. The average number of
babies on any given day is assumed to be 10, and the actual number $n$
realized is assumed to follow a Poisson distribution. The hospital/staff is
calibrated to handle 12 babies, and any number $n>12$ creates tension. We call
$X=\max(n-12,0)$ the level of tension, $Y$ the event as to whether there is
full moon ($Y=1)$ or no full moon $(Y=0)$. A signal is a pair $x=(X,Y)$ and
for each $x$ we can compute the ratio $f(x|1)/f(x|2),$ hence whether the
signal is evidence for $\theta=1$ or $\theta=2$, as well as the strength of
the evidence.\footnote{We assume that a full moon lasts 3 days out of 30.
Under state 2, the expected number of deliveries is $\alpha=10$ independantly
of the moon phase. \textit{Under state 1}, we denote by $\alpha_{1}$
(respectively $\alpha_{0}$) the expected number of deliveries on a full moon
day (respectively on other days). We have $\alpha_{1}=1.2$ $\alpha_{0}$ and
$\alpha_{1}+9\alpha_{0}=10\alpha$, so $\alpha_{1}=11.76$ and $\alpha_{0}%
=9.80$. We then use the Poisson distribution $P_{\alpha}(n)=e^{-\alpha}%
\alpha^{k}/k!$ to derive the tables. For example, for $X>0$, $l(X,0)=P_{\alpha
}(12+X)/P_{\alpha_{0}}(12+X)$ and $l(X,1)=P_{\alpha_{1}}(12+X)/P_{\alpha
}(12+X)$.} The signals that are evidence in favor of $\theta=2$ have the
following strength:
\[%
\begin{tabular}
[c]{|c|c|c|c|c|c|c|c|c|c|}\hline
{\small X,Y} & {\small 0,1} & {\small 8,0} & {\small 7,0} & {\small 6,0} &
{\small 5,0} & {\small 4,0} & {\small 3,0} & {\small 2,0} & {\small 1,0}%
\\\hline
${\small l}$ & {\small 1.31} & {\small 1.22} & {\small 1.20} & {\small 1.17} &
{\small 1.15} & {\small 1.13} & {\small 1.10} & {\small 1.08} & {\small 1.06}%
\\\hline
\end{tabular}
\ \ \
\]
and the strongest of these is $(0,1)$ (no tension on a full moon day). Signal
that are evidence in favor of $\theta=1$ have the following strength:%
\[%
\begin{tabular}
[c]{|c|c|c|c|c|c|c|c|c|c|}\hline
{\small X,Y} & {\small 0,0} & {\small 1,1} & {\small 2,1} & {\small 3,1} &
{\small 4,1} & {\small 5,1} & {\small 6,1} & {\small 7,1} & {\small 8,1}%
\\\hline
${\small l}$ & {\small 1.02} & {\small 1.41} & {\small 1.67} & {\small 1.96} &
{\small 2.31} & {\small 2.71} & {\small 3.19} & {\small 3.76} & {\small 4.42}%
\\\hline
\end{tabular}
\ \ \
\]
so, apart from signal (0,0), which is almost uninformative, they are stronger
compared to those in favor of $\theta=2$. In Figure \ref{fig10} below, we plot
the distributions over signals conditional on each state, and the induced
strength of evidence. In the Figure, the length of the horizontal segment
associated with signal $x$ indicates the probability of occurence of that
signal under $\theta=2)$.\footnote{For lisibility, we omit the most likely
signal $(0,0)$ and report the distributions conditional on the event
$x\neq(0,0)$.}

\begin{figure}[h]
\centering
\includegraphics[scale=0.65]{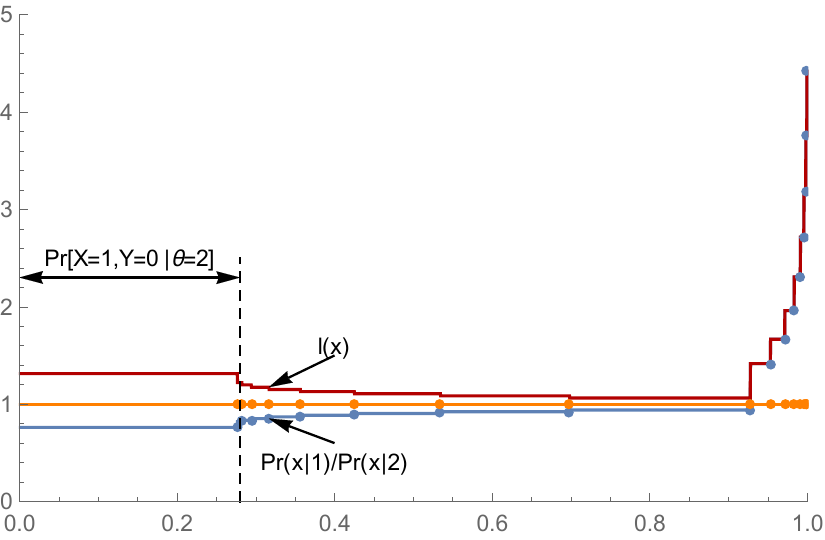}
\caption{Lunar effects: distributions and strength of evidence}%
\label{fig10}%
\end{figure}Under the simple mental processing considered earlier, and if
$\beta>0.31$, evidence in favor of $\theta=2$ is never processed, leading the
individual to believe in a lunar effect independently of the true
state.\footnote{Note that a smaller threshold would not make the problem
regular: for smaller values of the threshold $\beta$, evidence is
preponderantly in favor of $\theta=2$, \textit{independently of the underlying
state.}}

\paragraph{\textit{Illusory correlation and pattern identification. \ }}

Superstitions and other folk beliefs can also be interpreted as an instance of
illusory correlation (\cite{chapman67}) between two events, or more generally,
the illusory identification of a pattern in the environment.

Formally, one can think of a pattern as a sequence of two events $P$ and $C$
where $P$ is a premise and $C$ a consequence. The issue is whether the premise
makes the consequence more likely. Sometimes $PC$ is observed, but at other
times, $\overline{P}C$ or $P\overline{C}$ or $\overline{P}\overline{C}$ can
also be observed.\footnote{We denote by $\overline{P}$ the absence of a
premise and $\overline{C}$ the absence of the consequence.} As we explain
below, when $P$ and $C$ are both rare events, the only event which has
significant strength is the observation of $PC$ and it favors the theory that
an influence exists. To see this, assume that under state $\theta=1$, an
influence exits, while under state 2 it does not:
\begin{align*}
\Pr(C\;|\;P,\theta\left.  =\right.  1)  &  =\alpha\Pr(C\;|\;\overline
{P},\theta\left.  =\right.  1)\text{ and }\\
\Pr(C\;|\;P,\theta\left.  =\right.  2)  &  =\Pr(C\;|\;\overline{P}%
,\theta\left.  =\right.  2)
\end{align*}
with $\alpha>1$. Denote by $r=\Pr(P)$ and $q=\Pr(C)$. Letting $q_{1}%
=\Pr(C\;|\;P,\theta=1)$ and $\overline{q}_{1}=\Pr(C\;|\;\overline{P},\theta=1)
$, we have $rq_{1}+(1-r)\overline{q}_{1}=q$, implying
\[
\overline{q}_{1}=\frac{q}{1+(\alpha-1)r}<q<q_{1}=\frac{\alpha q}%
{1+(\alpha-1)r}.
\]
This gives us the direction and strength of evidence $(\overline{\theta},l)$
for each signal $x\in\{PC,\overline{P}C,P\overline{C},\overline{P}\overline
{C}\}$:%
\[%
\begin{tabular}
[c]{|c|c|c|c|c|}\hline
& $\overline{P}C$ & $P\overline{C}$ & $\overline{P}\overline{C}$ &
$PC$\\\hline
$\overline{\theta}$ & 2 & 2 & 1 & 1\\\hline
$l$ & $\frac{q}{\overline{q}_{1}}$ & $\frac{1-q}{1-q_{1}}$ & $\frac
{1-\overline{q}_{1}}{1-q}$ & $\frac{q_{1}}{q}$\\\hline
\end{tabular}
\]
When $P$ and $C$ are both rare events (i.e., $q$ and $r$ small), $\overline
{q}_{1}$ and $q_{1}$ are small as well, so $\frac{1-q}{1-q_{1}}$ and
$\frac{1-\overline{q}_{1}}{1-q}$ are close to 1. In addition, $\frac
{q}{\overline{q}_{1}}=1+(\alpha-1)r$ remains close to 1 while $q_{1}%
/q=\frac{\alpha}{1+(\alpha-1)r}$ is comparable to $\alpha$. It follows that
the strength of $PC$ is significantly higher than that of all other signals,
and it favors theory $\theta=1$. Of course, a proper weighting of all evidence
along with the Bayesian aggregation rule should eventually lead individuals to
avoid erroneous beliefs. However, under simple processing and if weak evidence
is ignored, the mental system inevitably points towards high belief states
(for which individuals are inclined to think that influence exists ($\theta=1$).

\subsection{Framing and pooling}

For a Bayesian, the frequency with which updating occurs is irrelevant. Nor
does it matter whether signals are pooled or not: to the extent that the
distributions $f(.\mid\theta)$ over signals are statistically distinguishable,
a Bayesian learns the correct state. Which alternative $\theta^{\prime}$ is
pitted against the true state $\theta$ does not matter either. So long as
$\theta$ is the true state, a Bayesian will learn it. Under our simple
belief-formation assumption, the frequency of updating, how signals are pooled
and how the problem is framed may all affect long-run beliefs, because this
affects the strength of evidence in favor of each underlying state, hence
eventually which signals are processed and which are not.\footnote{More
genrally, any prior views about signal generation may affect the perceived
strength of evidence, hence may distort beliefs.} This may explain the
persistence of erroneous beliefs, as well as the persistence of disagreement
among people despite the presence of common signals.

\textit{Batch processing.} Assume that instead of processing signals
$x_{1},..,x_{n},...$ sequentially (and updating the mental state after each
one), signals are processed by batches of $J$ signals, say $X_{1}%
=(x_{1},..,x_{J})$, $X_{2}=(x_{J+1},..,x_{2J})$ etc.... For any given problem,
if $J$ is sufficiently large, then by the law of large number, most batches
generated under $\theta$ are strong evidence in favor of $\theta$, hence the
problem becomes regular even if it was not regular under frequent processing.
Dealing with batches of signals may of course be cognitively more demanding,
but to the extent that the agent categorizes batches correctly (i.e.,
$\widetilde{\theta}=\overline{\theta})$ or with some errors but without
introducing systematic biases, biased beliefs can be avoided. Conversely, this
illustrates that frequent updating may contribute to biased beliefs.

\textit{Pooling signals. }Another source of bias may come from the way signals
are pooled. In our lunar effect example, the no-tension event $X=0$ pools all
events where the number of deliveries $n$ is below or equal to 12. If these
events were not pooled, and if a low $n$ were processed on a full moon, then
events $(n,1)$ with low $n$ could be processed, and this would be reasonably
strong evidence in favor of the no-lunar effect hypothesis. So the way signals
are pooled affects their strength, and, under our simple belief-formation
rules, this affects long-run beliefs.

In the case of deliveries, we chose to pool all realizations $n\leq12$ into
$X=0$ (no tension). One justification could be that observing low $n$ is
difficult, as there are always programmed deliveries that makes the number of
unprogrammed ones difficult to observe.

But there may be other reasons. People tend to be looking for explanations for
unlikely events that they observe, and the act of looking for explanations may
be event dependent, hence may affect which signals are actually recorded
and/or processed. For example, imagine that one does not even wonder whether
there is a full moon (or an absence of full moon) when there\ is no tension
(as one does not a priori see the full moon as a plausible cause for lack of
tension). This would mean that signals $(X,Y)=(0,1)$ and $(0,0)$ are pooled
into $X=0$. For a Bayesian that understands this selection process, this is
not an issue, as $X$ remains (weakly) informative, and in the long run, she
would correctly assess that a lunar effect does not exist if there is none.
For our less sophisticated agent, signal $X=0$ has only very weak informative
value, hence likely falls under the radar.\footnote{In the same vein, another
interpretation is that all events that seem irrelevant are pooled with truly
irrelevant ones. Again, for a Bayesian, this makes the ``irrelevant pool" not
so irrelevant, but it affects the long run mental state and beliefs of agents
that ignore these events.}


\textit{Framing. }To illustrate as simply as possible the effect of framing,
assume that we draw a biased coin with a probability $\alpha_{1}=0.7$ of
showing a $T$ail (rather than a $H$ead). Imagine that each signal is a draw
and that we test this theory ($\theta=1$, i.e., $\alpha_{1}=0.7$) against the
alternative theory $\theta=2$ with $\alpha_{2}=0.3$. Then the event $T$ is
evidence for $\theta=1$, while $H$ is evidence for $\theta=2$, and mental
states therefore point towards the correct underlying state. In contrast, if
the alternative theory is $\alpha_{2}=0.8$, the agent will more frequently see
evidence for $\theta=2$ than against it, and could thus erroneously conclude
that $\theta=2$ is the more likely state.

If a signal processed is a \textit{sequence of draw} rather than a single
draw, the issue is alleviated (because as explained above batch processing
tends to make the problem regular). But the issue persists so long as the
sequence remains small enough, with a key role played by the censoring
threshold $\beta$ in shaping the long-run distribution over mental states
hence beliefs.

\section{Discussion and Extensions}

\subsection{Fewer signals\label{SectionFewerSignals}}

Our analysis has so far assumed an arbitrarily large number of signals. We
discuss below the consequence of individuals only processing a limited number
of signals.

Formally, given $p$, call $\phi_{\theta}^{N,p}$ the distribution over mental
states when $\theta$ is the underlying state and $N$ the number of signals
processed, and let $\Phi_{\theta}^{N,p}(k)=%
{\textstyle\sum\nolimits_{s\geq k}}
\phi_{\theta}^{N,p}(s)$. If the decision maker chooses action 1 when her
mental state is at least $k$, she obtains
\[
W^{N}(p,k)\equiv\pi(1-\gamma)\Phi_{1}^{N,p}(k)+(1-\pi)\gamma(1-\Phi_{2}%
^{N,p}(k))
\]
When following $\sigma^{d}$ with a realized prior $\widetilde{\rho}$, her
decision rule calls for choosing action 1 when her mental state is at least
$k_{d,\widetilde{\rho}}\equiv\min\{s,\widetilde{\rho}s^{d}\geq\Gamma\}$, so
the expected welfare at $p$ is:%
\[
W_{p}^{N}(\sigma^{d})\equiv EW^{N}(p,k_{d,\widetilde{\rho}})
\]
The number of signals processed thus affects welfare insofar as $N$ affects
the cumulatives $\Phi_{\theta}^{N,p}$. With a limited number of mental states,
convergence is fast, and the effect is therefore limited. With 5 states and 10
signals for example, the maximum difference between cumulatives $\Phi_{\theta
}^{N,p}(k)$ and $\Phi_{\theta}^{\infty,p}(k)$ is at most equal to $4\%$
\textit{uniformly over the transition probabilities }$p$.

\begin{figure}[h]
\centering
\begin{minipage}{0.6\textwidth}
\centering
\includegraphics[scale=0.47]{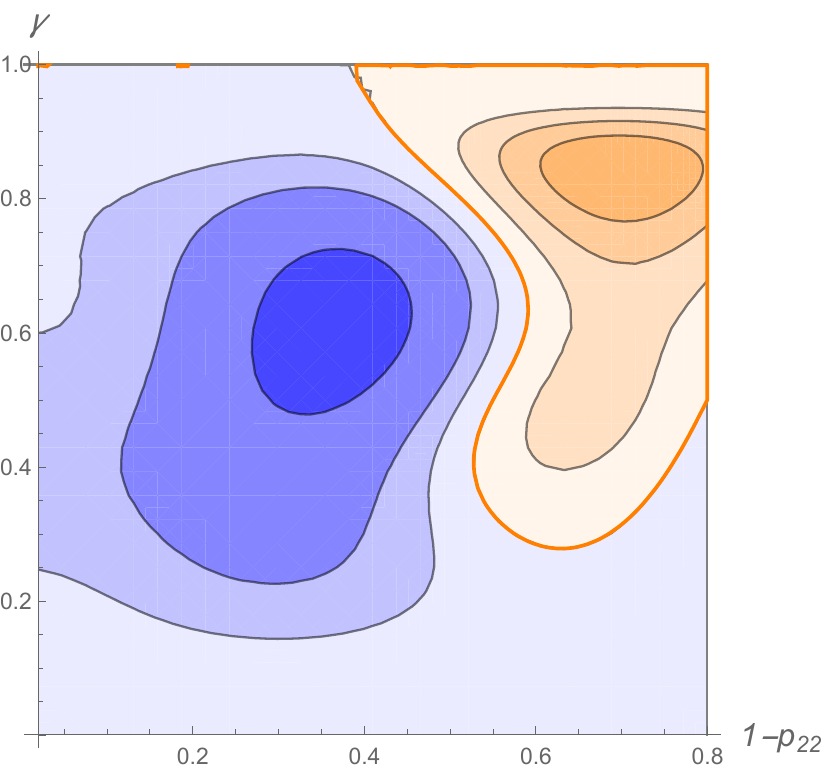}
\end{minipage}\begin{minipage}{0.1\textwidth}
\centering
	 \includegraphics[scale=0.38]{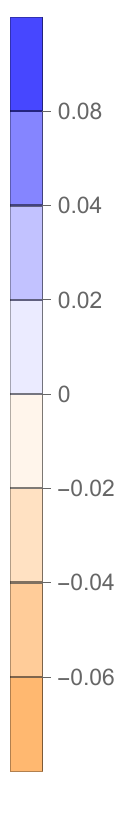}
\end{minipage}\hfill\caption{Welfare consequences of raising $N$}%
\label{fig11}%
\end{figure}

Regarding the direction of change, i.e., whether processing more signals
improves or hurts welfare, the answer depends on the locus of $p$. Figure
\ref{fig11} below reports the magnitude and sign of the ratio $\frac
{W_{p}^{\infty}-W_{p}^{10}}{\underline{W}}$, for a fixed $d=3$. For regular
problems, more signals help because they tend to increase the probability to
end up in an extreme state (a high state under $\theta=1$ and a low state
under $\theta=2)$. For non-regular problems however, getting more signals may
increase the loss. When $p_{11}>1/2$ and $p_{22}<1/2$ for example, processing
more signals tends to generate positive mental states independently of the
underlying state: the agent's decision is more subject to the mental system's
bias towards $\theta=1$,\ hence the higher losses when $\gamma>1/2$ (i.e.,
when on average $a=2$ is a better decision).


\subsection{More belief/mental states}

With more mental states, the mental system is potentially more efficient in
aggregating information. For example, in the Bayesian benchmark (where $d_{p}$
and $\Lambda_{p}$ can be both adjusted to the $p$ induced by censoring and to
the number of mental states), the set of problems for which the mental system
helps mostly expends, with a significant percentage gain for many problems. We
illustrate this below with a change from 5 to 7 states. We report the welfare
gain ratio $r=\Delta W/\underline{W}$ as a function of $p_{22}$ and $\gamma$,
assuming noisy priors. The left figure is the Bayesian benchmark (with mostly
gains). We keep $\sigma^{d}$ fixed with $d=3$ in the right figure:

\begin{figure}[h]
\centering
\begin{minipage}{0.35\textwidth}
\centering
\includegraphics[scale=0.47]{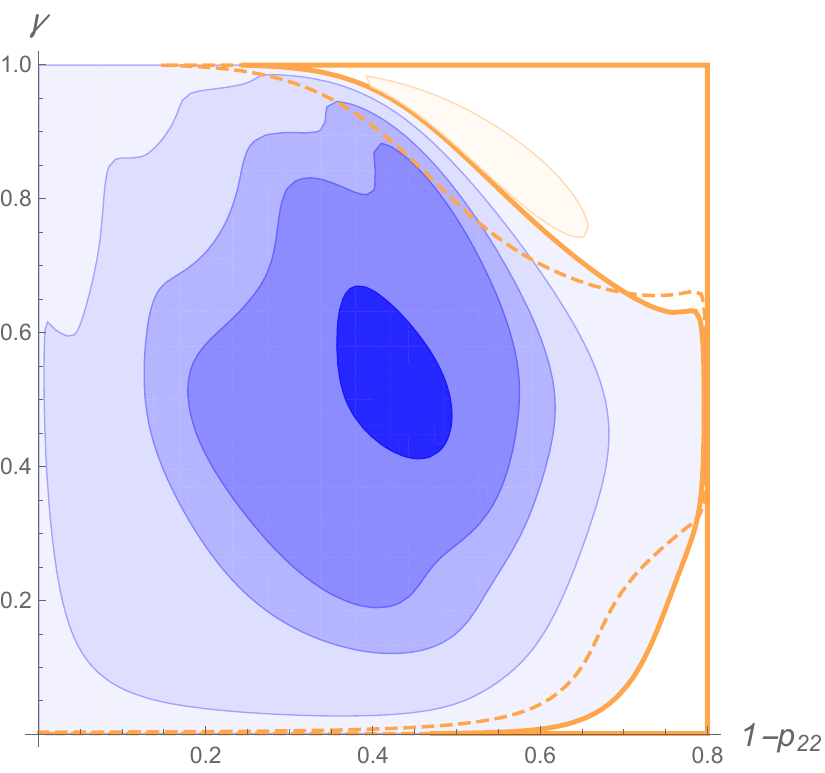}
\subcaption{Bayesian case}
\end{minipage}\begin{minipage}{0.1\textwidth}
\centering
	 \includegraphics[scale=0.6]{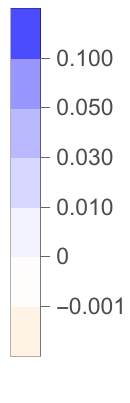}
\end{minipage}\hfill\centering
\begin{minipage}{0.35\textwidth}
\centering
\includegraphics[scale=0.47]{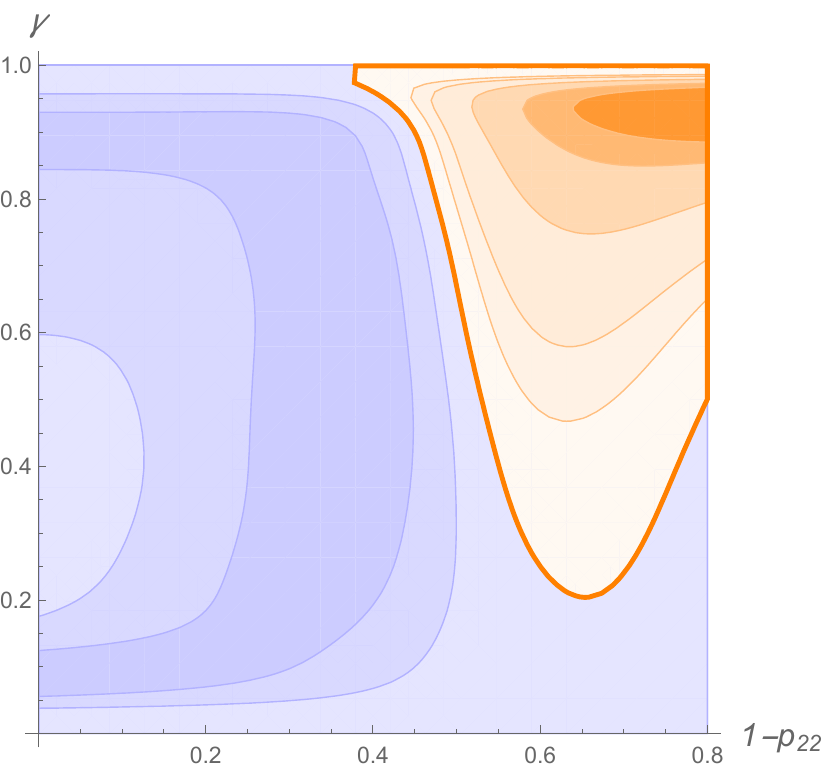}
\subcaption{fixed $\sigma^d$, with $d=3$}
\end{minipage}\begin{minipage}{0.1\textwidth}
\centering
	 \includegraphics[scale=0.6]{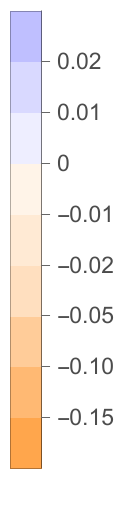}
\end{minipage}\hfill\caption{Welfare consequences of raising the number of
mental states}%
\label{fig12}%
\end{figure}

The right figure illustrates that increasing the number of states has mixed
consequences. The trade-off is similar to the one discussed in previous
sections: a higher number of states increases welfare for (most) regular
problems, but it diminishes welfare for some irregular problems.

This suggests that \textit{even in the absence of costs associated with
maintaining a larger number of mental states}, there may be a cost associated
with the more complex mental system. It performs slightly better on many
problems, but significantly worse for some.

Of course, the individual could adjust $d$ downward when he has 7 mental
states rather than 5. By reducing $d$ down to $d^{2/3},$ the spread in beliefs
remains the same whether he has 5 or 7 states: this would limit the gains in
the region of regular problems, but avoid the adverse consequence of keeping a
large $d$ in case the problem is irregular.

Nevertheless, to the extent that $d$ is an instrument that one finds difficult
to adjust, \textit{limiting the number of mental states }can be viewed as an
alternative instrument for reducing the risk of falling prey to mental
processing biases.

Thus, beyond the classic motive that mental states are scarce cognitive
resources, we suggest here an alternative motive for reducing the number of
mental states: with uncertainty about the data-generating process, too many
states may actually hurt welfare.

\subsection{Richer perceptions and more complex mental systems}

We assumed coarse perceptions $\widetilde{\theta}\in\{0,1,2\}$. We briefly
discuss an extension where perceptions include a noisy estimate $\widetilde{l}%
$ of likelihoods. Formally, this means that the agent now processes a sequence
$\widetilde{H}$ of perceptions each of the form $\widetilde{h}%
=(\widetilde{\theta},\widetilde{l})$. One issue is how these perceptions
should be aggregated and eventually generate beliefs. A natural extension of
our belief-formation model would be to keep the same belief-formation rule
($\widehat{\rho}=\widetilde{\rho}d^{s})$, but allow for mental-state changes
tuned to the perceived strength of the evidence, for example allowing for
\textit{two-step moves} (if feasible) in case $\widetilde{l}>d^{3/2}$.

With sufficiently accurate perception of strength, and despite the coarse
moves assumed (one or two steps), this type of mental processing is likely to
be helpful for some problems, as the mental-state transitions are now better
tuned to the real informativeness of the signals being processed. There are
several caveats however:

(i) Estimating the strength of evidence seems much more demanding than
estimating the direction of evidence

(ii) Even if estimates are correct, the issue we raised remains: if the agent
is unable to perceive correctly the resulting ex ante balance between
(properly weighted) confirming and disconfirming evidence, beliefs will be biased.

(iii) With sufficiently noisy perception of strength, the process gives rise
to random moves of $0,1$ or $2$ steps, and this more complex mental processing
may actually deteriorate welfare compared to the simple mental processing we
discussed (See \cite{compte10} for an example along those lines).\footnote{The
discussion, as well as the previous one about the number of mental states,
echoes the classic observation that complexity may come with lower fitness.}

Finally, we note that instead of the simple extension proposed here, some
might argue in favor of Bayesian-like belief-formation rules. For example, one
could consider a continuum of mental states $s$ and, starting from $s=1$, an
updating rule for mental states that sets (for any signal processed)
$s^{\prime}=s\widetilde{l}$ if $\widetilde{\theta}=1$ and $s^{\prime
}=s/\widetilde{l}$ if $\widetilde{\theta}=2$. After processing many signal,
this would lead to a mental state $\widehat{s}$,\footnote{With $d=1$,
$\widehat{\rho}=\widetilde{\rho}\widehat{s}=\widetilde{\rho}%
{\textstyle\prod\limits_{\widetilde{h},\widetilde{\theta}=1}}
\widetilde{l}/%
{\textstyle\prod\limits_{\widetilde{h},\widetilde{\theta}=2}}
\widetilde{l}$, which corresponds to the subjective Bayesian posterior.} and
the belief-formation rule $\widehat{\rho}=\widetilde{\rho}\widehat{s}^{d}$
would have a Bayesian-flavor: with $d=1$, the belief-formation rule would
induce subjective Bayesian updating (based on possibly biased perceptions of
strenght). A fully Bayesian agent would find the correct posterior conditional
on $\widehat{s}$, given the joint distribution on $(\widehat{s},\theta)$
induced by the distribution signals for the current problem and the
information processing constraints assumed.

Our approach advocates an intermediate route, which allows the agent to get
perceptions in accord with the particular problem faced $(\widetilde{h}$ is
correlated with $(\overline{\theta},l)$), but at the same time prevents a fine
tuning of posterior beliefs to the characteristics of that problem (i.e., to
the joint distribution over $(\widehat{s},\theta)$).

\subsection{More states of world}

We have considered an agent attempting to discriminate between only two
underlying states of the world. What if the agent attempts to discriminate
between more than two states?
We suggest below a simple extension of our model, illustrating that a
one-dimensional belief formation rule remains feasible and would perform well
under some conditions; but also highliting that
even if weak information is \textit{not} ignored, a bias towards theories that
generate strong evidence to likely to arise, in particular theories that
``see" patterns that happen to fit the data very well, some of the
time.\footnote{This observation is closely related to \cite{levy21}, who
assume that like us that people favor explanations that maximize the
likelihood of the data. One difference with \cite{levy21} is that we assume
that the data in processed in small pieces, a signal at a time, reinforcing
the tendency to select such more extreme theories.}

Formally, regarding the processing of signals, we proceed as before. We
associate to each signal $x$ a direction and strength of evidence. That is, we
again define $\overline{\theta}$ as the underlying state that best fits the
signal $x$, and $l$ as the degree to which $\overline{\theta}$ fits better the
data $x$ against all other alternatives, i.e.,\footnote{Note that a Bayesian
would need to keep track of \textbf{all} ratios $f(x\mid\overline{\theta
})/f(x\mid\theta)$, rather than the minimal value of these thresholds.}
\[
\overline{\theta}=\arg\max_{\theta}f(x\mid\theta)\text{ and }l=f(x\mid
\overline{\theta})/\max_{\theta\neq\overline{\theta}}f(x\mid\theta)
\]
Regarding the mental system, we assume $3K+1$ mental states, with states
labelled as $0$ or $(i,k)$ with $i\in\{1,2,3\}$ and $k\in\{1,..,K\}$. We
interpret a mental state $s=(i,k)$ as indicating overall evidence pointing
towards state $\theta=i$, to a degree $k$. Accordingly, starting from $s=0$,
we assume that when the agent processes a signal in favor of $\theta=i$, his
state moves up one step on the $i$-ladder if $s=$ $0$ or $(i,k)$ with $k<K$,
and otherwise (i.e., if on a $j$-ladder with $j\neq i$) moves down one step
(possibly reverting to $s=0$).

Regarding how beliefs are formed, let $\rho_{ij}=\Pr(i)/\Pr(j)$ denote the
prior likelihood. We assume that when in state $(i,k)$, the posterior
likelihood of $i$ against $j$ is:
\[
\rho_{ij}d^{k}
\]
In other words, the mental state can reinforce a belief in one particular
state, but it cannot modify the relative probabilities of low probability states.

Let again $p_{\overline{\theta}\theta}=\Pr(\overline{\theta}\mid\theta) $. It
should be clear that if $p_{kk}>1/2$ for all $k$, then the mental system,
however limited, improves welfare as it creates a positive correlation between
the underlying state $\theta$ and the set of mental states $\{(\theta
,k)\}_{k\geq1}$. But it is also easy to come up with problems for which
evidence for some theory is always inexistent, independently of the underlying state.

For example, consider an agent receiving a sequence $x=(x_{1},...,x_{5})$ of 6
draws of 1's and $0$'s possibly autocorrelated. Let $\rho=\Pr(x_{m+1}=x_{m})$
and assuming that possible values of $\rho$ are $2/3$ ($\theta=1$)$,1/3$
($\theta=2$) and $0.5$ ($\theta=3)$. The following table gathers the pairs
$(\overline{\theta},l)$ for each sequence received as a function of the number
$n$ of reversals (i.e., $x_{m+1}\neq x_{m}$).
\[%
\begin{tabular}
[c]{|c|c|c|c|c|c|c|}\hline
${\small n}$ & ${\small 0}$ & ${\small 1}$ & ${\small 2}$ & ${\small 3}$ &
${\small 4}$ & ${\small 5}$\\\hline
$\overline{\theta}$ & {\small 1} & {\small 1} & {\small 1} & {\small 2} &
{\small 2} & {\small 2}\\\hline
${\small l}$ & {\small 4.2} & {\small 2.1} & {\small 1.05} & {\small 1.05} &
{\small 2.1} & {\small 4.2}\\\hline
$\Pr{\small (x\mid\theta=3)}$ & {\small 0.03} & {\small 0.16} & {\small 0.31}
& {\small 0.31} & {\small 0.16} & {\small 0.03}\\\hline
\end{tabular}
\]
With sequences of 6 draws, there is no sequence that provides clear evidence
in favor of independence, and the agent is lead to believe either in positive
or negative autocorrelation even when there is no autocorrelation. With
sequences of limited length, there is no sequence that is an obvious
representative of an independent sequence of draws.

As the number of elements in a sequence increases, evidence in favor of
independence surfaces for some draws and become frequent, but even for
sequence of 10 draws, the evidence tends to be weak compared to evidence in
favor of other states. When $x$ consists of a sequence of 10 draws, we have:%
\[%
\begin{tabular}
[c]{|c|c|c|c|c|c|c|c|c|c|c|}\hline
{\small n} & {\small 0} & {\small 1} & {\small 2} & {\small 3} & {\small 4} &
{\small 5} & {\small 6} & {\small 7} & {\small 8} & {\small 9}\\\hline
$\overline{\theta}$ & {\small 1} & {\small 1} & {\small 1} & {\small 1} &
{\small 3} & {\small 3} & {\small 2} & {\small 2} & {\small 2} &
{\small 2}\\\hline
${\small l}$ & {\small 13} & {\small 6.7} & {\small 3.3} & {\small 1.7} &
{\small 1.2} & {\small 1.2} & {\small 1.7} & {\small 3.3} & {\small 6.7} &
{\small 13}\\\hline
$\Pr{\small (x\mid\theta=3)}$ & {\small 0.002} & {\small 0.02} & {\small 0.07}
& {\small 0.16} & {\small 0.25} & {\small 0.25} & {\small 0.16} &
{\small 0.07} & {\small 0.02} & {\small 0.002}\\\hline
\end{tabular}
\]

When $\theta=3$, evidence in favor of $3$ is frequent, but never quite
striking, unlike evidence for other states of the world. Evidence in favor of
more extreme states is more striking.

\subsection{A motive for stake-contingent skepticism.}

To conclude this Section, we comment on the classic separation between beliefs
and preferences inherited from Ramsey and Savage. In a Bayesian model, agents
form beliefs based on signals, independently of the stakes involved. In our
model, the issues become intertwined: we endogenize belief formation (through
$d$ for example), and when the stakes $\gamma$ are larger, the agent has
incentives to decrease $d$, i.e., use more caution in forming beliefs.

To see why, we fix again $d=3$ and compare the magnitudes of gains and losses
for $\gamma=0.6$ and $\gamma=0.75$ across all possible $p$:

\begin{figure}[h]
\centering
\begin{minipage}{0.35\textwidth}
\centering
\includegraphics[scale=0.45]{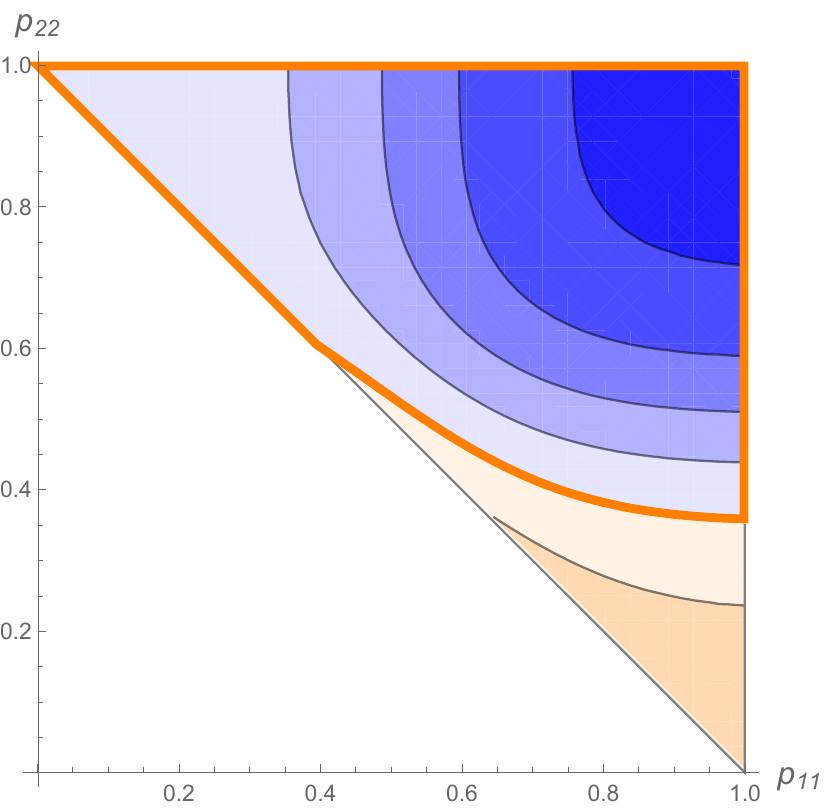}
\subcaption{$\gamma=0.6$}
\end{minipage}
\begin{minipage}{0.1\textwidth}
\centering
\includegraphics[scale=0.45]{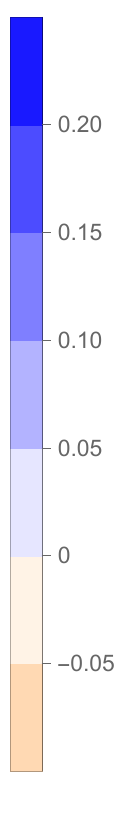}
\end{minipage}\hfill\begin{minipage}{0.35\textwidth}
\centering
	 \includegraphics[scale=0.45]{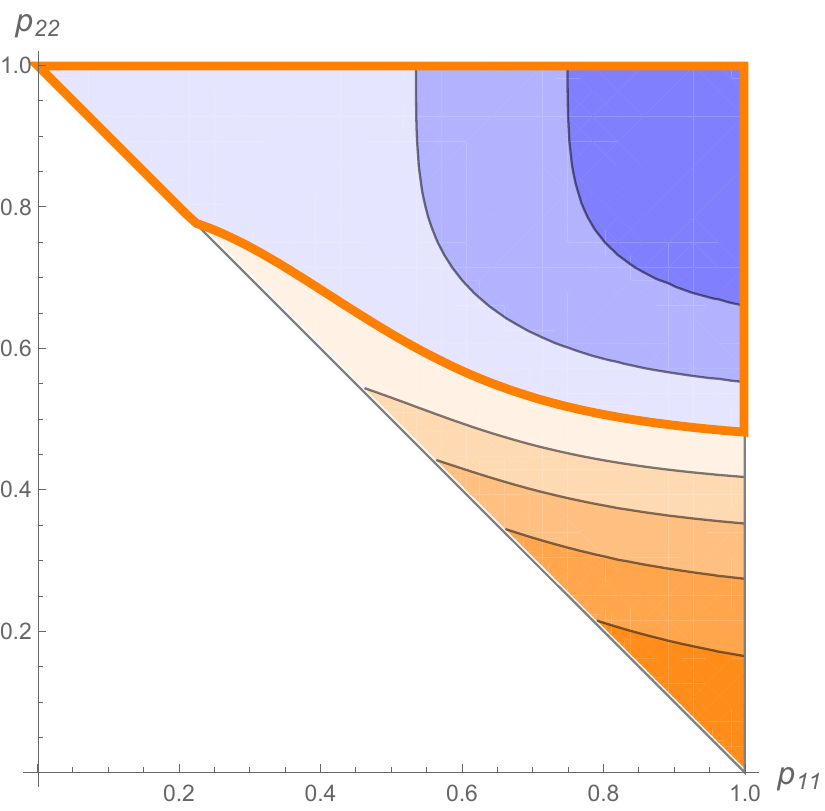}
\subcaption{$\gamma=0.75$}
\end{minipage}
\begin{minipage}{0.1\textwidth}
\centering
\includegraphics[scale=0.45]{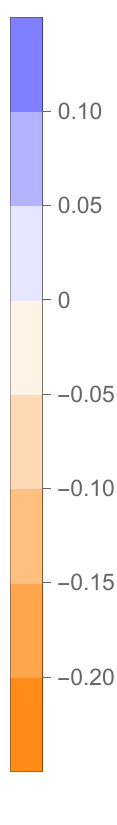}
\end{minipage}
\caption{Welfare gains for different stakes $\gamma$}%
\label{fig13}%
\end{figure}Incentives to set $d$ depends on the distribution over problems
$p$ faced, but it should be clear from the figure that when $\gamma$ is high,
losses become preponderant, thus providing the agent with incentives to
decrease $d$ and give a more prominent role to priors.

Intuitively, the agent faces two kinds of problems: some for which evidence is
somewhat balanced (i.e., $\Lambda_{p}$ is not too far from $1$) and some for
which evidence on average points in a given direction, say $\widehat{\theta
}=1,$ independently of the state $\theta.$ When the agent ends up in a high
mental state, this is evidence in favor of $\theta=1$ for the first set of
problems, but this is not the case for the second set of problems.

When $\Gamma/\rho=1$, relying on priors gives the lowest possible welfare, so,
for the second set of problems, being erroneously influenced by the mental
state is not costly. When $\Gamma/\rho$ is large however, this influence is
costly: for the second set of problems, the agent is mislead into thinking
that $\theta=1$ while he would have been better off following priors.

In other words, for asymmetric-stake cases, the agent may benefit from being
more cautious and exert some stake-contingent skepticism, which can be done by
reducing $d$ when stakes are higher.

\subsection{Further comments}

\textbf{Endogenous classification.} We defined $\widetilde{\theta}\in\{1,2\}$
to indicate whether a given signal $x$ is perceived as evidence for $\theta=1$
or $\theta=2$ and $\widetilde{\theta}=0$ when the signal is not processed. In
doing this categorization, the agent is assumed to be using a
\textit{classification heuristic} based on likelihood ratios (the ratio of the
likelihood of $x$ under $\theta$ and of the likelihood of $x$ against the
alternative hypothesis). Through the choice of $\beta$, the agent ensures that
only sufficiently informative signals are processed.

So by endogenizing $\beta$, we endogenize the classification heuristic used by
the agent. Note that this classification is \textit{context dependent}, in the
sense that it depends on which underlying states $\theta$ are compared.

\textbf{Misspecified models.} Let us contrast our work with the literature
that explain biases through agents forming beliefs based on a misspecified (or
incomplete) model of the environment (See Spiegler (2020) for a review). In
our model, one can think of $(\theta,f)$ as an \textquotedblleft
extended\textquotedblright\ state of the world, describing both the underlying
state and the signal generating process for each possible state of the world.
The inferences that the agent draws from signals are tuned to $f$, so as far
as perceptions are concerned, the agent's model is not misspecified. But next,
we prevent the agent from tuning the belief formation rule to the induced
distribution over perceptions given $f$. She ends up using a belief-formation
rule of the form $\rho d^{s}$ as if the process generating perceptions (which
is characterized by the transition probabilities $p$ (given $f$ and $\beta)$)
induced a balanced mental system for which $\Lambda_{p}=1$ (which it is not
because in general, $\Lambda_{p}$ differs from $1$).

The logic by which we derive $d$ differs from that of a typical misspecified
model.\footnote{We think of it as a welfare maximizing heuristic within the
set of possible $d$'s, rather than the result of subjective Bayesian updating
under incorrect priors.} But for any given $d$ so derived, one may say that
the agent forms beliefs as if she had a misspecified prior over this
higher-level object, i.e., the joint distribution over mental states and
states of the world (induced by perceptions and the mental system): she forms
beliefs as she had the erroneous view that her mental system is balanced,
which is a reasonable hypothesis on average across problems, but not for the
specific one under consideration.

\section{Conclusion.}

We have modelled agents whose behavior is governed by two heuristics, one that
governs the classification of signals (through censoring $\beta$), and one
that governs caution in decision making (through the discrimination power $d$
that the agent assigns to the mental system). The agent adjusts these two
heuristics in the direction of welfare improvements computed on average over
the various discrimination problems that the agent faces, thus without being
able to adjust the two instruments $\beta$ and $d$ to the specific
data-generating process considered. The optimal heuristic can only be good on
average, and our analysis highlights the type of discrimination problems for
which biases are generated, as well as how pooling and framing can be used to
distort one's belief. A more systematic study of this last phenomenon, and of
how a strategic party could exploit it for persuasion purposes, deserves
further research.

Our model also tried to separate the inferences that people make from each
signal (calling them perceptions) from the long-run mental state that
aggregates these perceptions. We point out that the task of forming beliefs
based on mental states is a difficult one: while each inference may be correct
for the discrimination problem considered, understanding, as a Bayesian would,
the properties of the joint distribution over mental states and states of the
world for that problem is a challenge.

\section*{Appendix}

We start with preliminary notations and observations. For any $r$, we define
the distribution $f_{r}$ over mental states having the property that
$f_{r}(s)=rf_{r}(s-1)$, which implies
\begin{equation}
f_{r}(s)=r^{s}f_{r}(0)\text{ and }f_{r}(0)=1/\sum_{s=-K}^{K}r^{s}.
\label{Eqf0}%
\end{equation}

\textbf{Lemma 1:} When $N$ is arbitrarily large, under state $\theta$, the
long-run distribution over mental state converges to $f_{r}$ where
$r=\frac{p_{1\theta}}{p_{2\theta}}$

\textbf{Proof of Lemma 1.} For any state $s\neq K$ or $-K$, the long-run
distribution, say $\phi$, satisfies
\begin{align*}
\phi(s)  &  =q_{1\theta}\phi(s-1)+q_{2\theta}\phi(s-1)+q_{0\theta}%
\phi(s)\text{ for }s\neq K\text{ or}-K\\
\phi(K)  &  =q_{1\theta}\phi(K-1)+(q_{1\theta}+q_{0\theta})\phi(K)
\end{align*}
which implies%
\begin{align*}
\phi(s)  &  =p_{1\theta}\phi(s-1)+p_{2\theta}\phi(s+1)\text{ for }k\neq
K\text{ or}-K\\
\phi(K)  &  =p_{1\theta}(\phi(K-1)+\phi(K))
\end{align*}
The second equality yields $\phi(s)=r\phi(s-1)$ for $s=K$, and, for $s<K$, the
equality is then obtained by induction on $s$.$\blacksquare$

Lemma 1 implies that the long-run distribution over mental states under
$\theta$ depends only on the conditional transition probabilities $p$.
Throughout the Appendix, we denote by $\phi_{\theta}^{p}(s)$ this distribution.

\textbf{Proof of Proposition\ 1. }For a given $p$, define $\Lambda
^{p}(s)\equiv\frac{\phi_{1}^{p}(s)}{\phi_{2}^{p}(s)}$. For a given prior
$\rho=\pi/(1-\pi)$, $\rho\Lambda^{p}(s)$ corresponds to the Bayesian posterior
belief about the underlying state, and the belief-formation rule $\sigma
^{\ast}(\rho,s)=\rho\Lambda^{p}(s)$ (which calls for choosing $a=1$ when
$\sigma^{\ast}(\rho,s)\geq\frac{\gamma}{1-\gamma}=\Gamma$) achieves the
maximum feasible welfare:\footnote{Summing over all $(\theta,s)$, expected
welfare is $W=%
{\textstyle\sum\nolimits_{s\in S}}
(\pi\phi_{1}^{p}(s)g(a(s),1)+(1-\pi)\phi_{2}^{p}(s)g(a(s),2)).$The agent's
welfare is maximum when for each $s$, the agent chooses $a=1$ when
$(1-\gamma)\pi\phi_{1}^{p}(s)\geq\gamma(1-\pi)\phi_{2}^{p}(s)$.}
\[
\overline{W}(p)=%
{\textstyle\sum\nolimits_{s\in S}}
\max((1-\gamma)\pi\phi_{1}^{p}(s),\gamma(1-\pi)\phi_{2}^{p}(s)).
\]
From Lemma 1, $\phi_{\theta}^{p}\equiv f_{r_{\theta}}$ with $r_{\theta}%
=\frac{p_{1\theta}}{p_{2\theta}}$, so $\Lambda^{p}(s)=\Lambda_{p}d_{p}^{s}$
where $d_{p}=r_{1}/r_{2}$ and $\Lambda_{p}=\frac{f_{r_{1}}(0)}{f_{r_{2}}(0)}%
$.It follows that when $\widetilde{\rho}=\rho$, $\sigma_{p}^{\ast}$ coincides
with $\sigma^{\ast}$ and thus achieves maximum possible welfare.$\blacksquare$

\textbf{Proof of Proposition 2}: Formally, $\Delta^{d}(p)$ can be expressed
as
\begin{align*}
\Delta^{d}(p)  &  =%
{\textstyle\sum\nolimits_{s}}
\psi^{p}(s)J^{d}(s)\text{, where}\\
\psi^{p}(s)  &  =(1-\gamma)\pi\phi_{1}^{p}(s)-\gamma(1-\pi)\phi_{2}%
^{p}(s)\text{ and }\\
J^{d}(s)  &  =\Pr(\mu\geq\Gamma/(\rho d^{s}))-\Pr(\mu\geq\Gamma/\rho).
\end{align*}
For $d>1$, $J^{d}(s)$ has the same sign as $s$, and by construction, for $p\in
B$, $\psi^{p}(s)$ also has the same sign as $s$, which proves the
proposition.\footnote{Note that when the number of mental states rises,
Condition (\ref{eqA}) becomes a more stringent one. The reason is that when
the number of mental states rises, being in state $0$ can become quite
informative if $p_{11}$ and $p_{22}$ differ (i.e., $\max(\Lambda_{p}%
,1/\Lambda_{p})$ becomes large).}$\blacksquare$

\textbf{Proof of Proposition 3. }The first order effect of raising $\beta$
above $0$ is to decrease all $q_{\widetilde{\theta}\theta}$ with
$\widetilde{\theta}\in\{1,2\}$ by some $x>0$. At $\beta=0$,
$p_{\widetilde{\theta}\theta}=q_{\widetilde{\theta}\theta}$, and we consider
the effect of $x$ on $p_{\theta\theta}$, $d_{\theta}$, $d$ and $\Lambda$,
expressing these as a function of $x$. Then we consider the marginal effect at
$x=0$. We have $p_{\theta\theta}(x)=\frac{p_{\theta\theta}-x}{1-2x}$, which
yields $\frac{\partial p_{\theta\theta}}{\partial x}|_{x=0}=2p_{\theta\theta
}-1,\ $which is positive when $p_{\theta\theta}>1/2$. Next we have
$d(x)=d_{1}(x)d_{2}(x)$ with $d_{\theta}(x)\equiv\frac{p_{\theta\theta}%
-x}{p_{\theta\theta^{\prime}}-x}$. Since $\frac{\partial d_{\theta}}{\partial
x}|_{x=0}=\frac{p_{\theta\theta}-p_{\theta\theta^{\prime}}}{(p_{\theta
\theta^{\prime}})^{2}}>0$ (because at $\beta=0$, $p_{\theta\theta}%
>p_{\theta\theta^{\prime}}$), we conclude that $d$ increases with $x$.

We now turn to $\Lambda=\frac{f_{r_{1}}(0)}{f_{1/r_{2}}(0)}$, where
$r_{\theta}=\frac{p_{\theta\theta}}{1-p_{\theta\theta}}$ and $f_{r}(0)$ is the
weight defined in Equation \ref{Eqf0}. We define $H(r)\equiv1/f_{r}(0)$ and
$h(r)\equiv r+1/r$. We have $H(r)=1+\sum_{s=1}^{K}h(r^{s})$, so $H(r)=H(1/r)$
and $H$ is strictly increasing in $r$ for $r>1$, with $H(1)=1+K$. Furthermore,
it is immediate to check that $H(r)=r^{-K}\sum_{k=0}^{2K+1}r^{s}=r^{-K}%
\frac{r^{2(K+1)}-1}{r-1}$. Letting $r_{\theta}\equiv\frac{p_{\theta\theta}%
}{1-p_{\theta\theta}}$ and $\overline{r}_{\theta}\equiv\max(r_{\theta
},1/r_{\theta})(>1)$, we obtain $\Lambda=\frac{H(\overline{r}_{2}%
)}{H(\overline{r}_{1})}$. So $\Lambda<1$ is equivalent to $\overline{r}%
_{1}>\overline{r}_{2}$.

We now investigate the effect of a marginal change in $x$. Note that
$\frac{\Lambda^{\prime}}{\Lambda}=\overline{r}_{2}^{\prime}\frac{H^{\prime
}(\overline{r}_{2})}{H(\overline{r}_{2})}-\overline{r}_{1}^{\prime}%
\frac{H^{\prime}(\overline{r}_{1})}{H(\overline{r}_{1})}$, so by checking that
$G(\overline{r})\equiv\overline{r}^{\prime}\frac{H^{\prime}(\overline{r}%
)}{H(\overline{r})}$ is increasing in $\overline{r}$ for $\overline{r}>1$, we
obtain $\Lambda^{\prime}<0$ whenever $\Lambda<1$, as desired.

We have $\overline{r}=\frac{p}{1-p}$ for some $p$, thus $\overline{r}%
(x)=\frac{p-x}{1-p-x}$, hence we get $\overline{r}^{\prime}=\frac
{2p-1}{(1-p)^{2}}=r^{2}-1$. Standard computations then show that $G^{\prime}$
has the same sign as
\[
g(r)\equiv(r^{2}+K(1+r^{2}))(r^{4+4K}-1)-(2+8K+4K^{2})(r^{2}-1)r^{2+2K}%
+2(r^{4+4K}-r^{2+4K})
\]
We have $g(1)=0$ and one can compute $g^{\prime}$, which has the same sign as
\[
g_{2}(r)=K(1+4K+2K)-2(1+K)(2+3K+K^{2})r^{2}+(2+K)(1+2K)r^{2+2K}%
+(1+K)(3+2K)r^{4+2K}.
\]
Again $g_{2}(1)=0$, and one easily checks that $g_{2}^{^{\prime}}>0$ for all
$r>1$, so $g_{2}$ is positive for $r>1$, so $g$ is positive as well for $r>1$,
so $G$ in increasing, which concludes the proof.$\blacksquare$

\textbf{Proof of Proposition 4}: the proof consists in defining $\overline
{\Lambda}_{p}=\Lambda_{p}(d_{p})^{K}$ and showing that the set $D=\{p,\frac
{\partial\overline{\Lambda}_{p}}{\partial\beta}<0\}$ is non-empty. Then,
although censoring increases $d_{p}$, the largest possible shift in posterior
beliefs decreases. Consider then a problem where $\Gamma/\rho$ is below but
close to $\overline{\Lambda}_{p}$. In the absence of censoring, the agent
chooses $a=1$ in the mental state $K$, and $a=2$ otherwise, so the mental
system improves welfare (over following priors). With censoring, all
posteriors remain below $\Gamma/\rho$, so the censoring makes the mental
system useless. We now show that $D$ is not empty.

With states labelled from 0 to $n$, $\phi_{1}(n)=r^{n}\phi_{1}(0)=\frac
{r_{1}^{n}}{\sum_{k=0}^{n}r_{1}^{k}}=\overline{H}(r_{1})$ where $\overline
{H}(r)=\frac{r^{n}(r-1)}{r^{n+1}-1}$. We thus hat $\overline{\Lambda}%
_{p}=\frac{\overline{H}(r_{1})}{\overline{H}(r_{2})}$ and $\frac
{\overline{\Lambda}_{p}^{\prime}}{\overline{\Lambda}_{p}}=r_{1}^{\prime}%
\frac{\overline{H}^{\prime}(r_{1})}{\overline{H}(r_{1})}-r_{2}^{\prime}%
\frac{\overline{H}^{\prime}(r_{2})}{\overline{H}(r_{2})}$. Define
$\overline{G}(r)=r^{\prime}\frac{\overline{H}^{\prime}(r)}{\overline{H}(r)}$.
We show below that $\overline{G}^{\prime}(r)$ for $r$ large enough. This will
imply that for $r_{1}>r_{2}>r$, $\overline{\Lambda}_{p}^{\prime}<1$. Recall
that $r^{\prime}=r^{2}-1$, so simple computations yield%
\[
\overline{G}(r)=\frac{(1+r)(n(1-r)+r(r^{n}-1))}{r(r^{n+1}-1)}%
\]
from which we obtain that for large $r$ the preponderant terms of
$\overline{G}^{\prime}$ are $-r^{2+2n}$ and $r^{3+n}(n^{2}+n)$. For any fixed
$n\geq2$, $\overline{G}^{\prime}$ is thus negative for large enough
$r$.$\blacksquare$

We now turn to Proposition 5. Let $\Phi(k,r)\equiv\frac{\sum_{s\geq k}r^{s}%
}{\sum_{s\in S}r^{s}}$. For any $p$, $\phi_{\theta}^{p}(s)$ depends only on
$r_{\theta}\equiv\frac{p_{1\theta}}{1-p_{1\theta}}$, with $\phi_{\theta}%
^{p}(s)=(r_{\theta})^{s}/\sum_{s\in S}(r_{\theta})^{s}$. We define
$\Phi_{\theta}^{p}(k)\equiv\sum_{s\geq k}\phi_{\theta}^{p}(s)$ as the
probability that the agent ends up in a mental state $s\geq k$ when the
underlying state is $\theta$ and transitions are described by $p$. By
construction $\Phi_{\theta}^{p}(k)=\Phi(k,r_{\theta})$. We have: \smallskip

\textbf{Lemma:} \textit{For any }$k>-K$\textit{, with }$k\leq K,$ $\Phi
(k,r)$\textit{\ strictly increases (decreases) with }$r$ \textit{if} $r>1$
$(r<1).$ \smallskip

\textbf{Proof of Lemma:} We relabel mental states from $n=0$ to $N=2K$ (so
that $n=s+K$) and let $f(n,r)\equiv\Phi(n-K,r)$. We have $f(n,r)=1-\frac
{r^{n}-1}{r^{N+1}-1}$. $\frac{\partial f}{\partial r}$ has the same sign as
$g(r)=(N+1-n)r^{N+1}-(N+1)r^{N+1-n}+n$. Since $g(1)=0$ and $g^{\prime
}(r)=(N+1)(N+1-n)r^{N-n}(r^{n}-1)>0$ for $r>1$ and $n\in\{1,..,N\}$, we
conclude that $g(r)>0$ for all $r>1$ and $n\in\{1,..,N\},$ so $f^{\prime}(r)$
is positive for all $r>1$ and $n\in\{1,..,N\}$ which concludes the
proof.$\blacksquare$


\textbf{Proof of Proposition 5: }Formally, consider any monotone strategy
$\sigma$ and any realization $\widetilde{\rho}$. Under $(\sigma
,\widetilde{\rho})$, the decision maker chooses action 1 if and only if the
mental state is high enough, say $s\geq k_{\sigma,\widetilde{\rho}}$, and the
welfare is given by $W(k_{\sigma,\widetilde{\rho}},p)$ where
\begin{equation}
W(k,p)\equiv\pi(1-\gamma)\Phi_{1}^{p}(k)+(1-\pi)\gamma(1-\Phi_{2}%
^{p}(k))\label{eqW}%
\end{equation}
where $\Phi_{\theta}^{p}(k)\equiv\sum_{s\geq k}\phi_{\theta}^{p}(s)$ is the
probability to end up in a mental state $s\geq k$ when the underlying state is
$\theta$.\footnote{This means that, over realizations of $\widetilde{\rho}$,
the agent obtains an expected welfare equal to $E_{\widetilde{\rho}%
}W(k_{\sigma,\widetilde{\rho}},p)$.} Now recalling that $\Phi_{1}^{p}%
(k)=\Phi(k,r_{1})$ and $\Phi_{2}^{p}(k)=\Phi(k,r_{2})$ with $r_{1}%
=\frac{p_{11}}{1-p_{11}}$ and $1/r_{2}=\frac{p_{22}}{1-p_{22}}$. For a regular
problem, both $r_{1}$ and $1/r_{2}$ increase with censoring, so we conclude
from the Lemma that with censoring, $\Phi_{1}^{p}(k)$ increases and $\Phi
_{2}^{p}(k)$ decreases. So welfare increases for any realization of
$\widetilde{\rho}$, hence it also increases on average over realizations of
$\widetilde{\rho}$.$\blacksquare$

\end{document}